\documentclass[12pt]{article}

\usepackage{amsmath}
\usepackage{epsfig}
\usepackage{amssymb}
\usepackage[colorlinks=true,
            linkcolor=blue,
            citecolor=blue]{hyperref}
\usepackage{bbold}
\usepackage{braket}
\input{epsf}
\usepackage[english]{babel}

\usepackage[numbers,sort&compress]{natbib}

\newcommand{\jet}{\text{jet}}

\def\to{\rightarrow}

\def\ep{\varepsilon}
\def\nn{\nonumber}
\def\beq{\begin{equation}}
\def\eeq{\end{equation}}
\def\beeq{\begin{eqnarray}}
\def\eeeq{\end{eqnarray}}
\def\beal{\begin{align}}
\def\eeal{\end{align}}

\begin{document}

\begin{titlepage}
\renewcommand{\thefootnote}{\fnsymbol{footnote}}
\par \vspace{10mm}

\begin{center}
\vspace*{10mm}
{\Large \bf
Hadron Fragmentation Inside Jets in Hadronic Collisions}
\end{center}

\par \vspace{2mm}
\begin{center}
{\bf Tom Kaufmann${}^{\,a}$,}
\hskip .2cm
{\bf Asmita Mukherjee${}^{\,b}$,}
\hskip .2cm
{\bf Werner Vogelsang${}^{\,a}$  }\\[5mm]
\vspace{5mm}
${}^{a}$ Institute for Theoretical Physics, T\"ubingen University, 
Auf der Morgenstelle 14, \\ 72076 T\"ubingen, Germany\\[2mm]
${}^{b}$ Department of Physics, Indian Institute of Technology Bombay, \\
Powai, Mumbai 400076, India
\end{center}


\vspace{9mm}
\begin{center} {\large \bf Abstract} \end{center}
We present an analytical next-to-leading order QCD calculation of the partonic cross sections
for the process $pp\rightarrow (\jet \,h)X$, for which a specific hadron is observed inside a fully 
reconstructed jet. In order to obtain the analytical results, we assume the jet to be relatively narrow.
We show that the results can be cast into a simple and systematic form based 
on suitable universal jet functions for the process. We confirm the validity of our calculation by 
comparing to previous results in the literature for which the next-to-leading order cross section 
was treated entirely numerically by Monte-Carlo integration techniques. We present phenomenological 
results for experiments at the LHC and at RHIC. These suggest that $pp\rightarrow (\jet \,h)X$ should 
enable very sensitive probes of fragmentation functions, especially of the one for gluons.

\end{titlepage}  

\setcounter{footnote}{2}
\renewcommand{\thefootnote}{\fnsymbol{footnote}}

\section{Introduction}

Final states produced at high transverse momentum ($p_T$), such as jets, single 
hadrons, or prompt photons, have long been regarded as sensitive and well-understood 
probes of short-distance QCD phenomena. Recently, a new ``hybrid'' type of 
high-$p_T$ jet/hadron observable has been proposed and 
explored theoretically~\cite{Procura:2009vm,Jain:2011xz,Procura:2011aq,Arleo:2013tya,Ritzmann:2014mka}.
It is defined by an identified specific hadron found inside a fully reconstructed jet, giving rise to a 
same-side hadron-jet momentum correlation. This correlation may for example be described in terms
of the variable $z_h\equiv p_T/p_T^{\,{\mathrm{jet}}}$, where $p_T$ and 
$p_T^{\,{\mathrm{jet}}}$ are the transverse momenta of the hadron and the jet, respectively. 
The production of identified hadrons in jets was first considered for the case of $e^+e^-$ 
annihilation~\cite{Procura:2009vm,Jain:2011xz,Procura:2011aq} and subsequently also for
$pp$ scattering~\cite{Arleo:2013tya}. Experimental studies have been pioneered in
$p\bar{p}\rightarrow (\jet \,h)X$ at the Tevatron~\cite{Abe:1990ii}. At the LHC, the
ATLAS~\cite{Aad:2011sc,ATLAS:2015mla} and CMS~\cite{Chatrchyan:2012gw} experiments 
have studied $pp\rightarrow (\jet \,h)X$,
and measurements are being carried out by ALICE~\cite{Hess:2014xba}. Measurements 
of the cross section (and, perhaps, spin asymmetries) should also be possible at RHIC.

There are several reasons why it is interesting to study the production of hadrons inside jets. 
Perhaps most importantly, the observable provides an alternative window on fragmentation functions~\cite{Arleo:2013tya}.
The latter, denoted here by $D_c^h(z,\mu)$, describe the formation of a hadron $h$ from a parent
parton $c=q,\bar{q},g$. The variable $z$ is the fraction of the parton's momentum transferred to the hadron, 
and $\mu$ denotes the factorization scale at which the fragmentation function is probed. Usually, fragmentation functions 
for a hadron $h$ are determined from the processes $e^+e^-\to hX$ or $ep\to e h X$. The power of these processes
lies in the fact that they essentially allow direct scans of the fragmentation functions as functions of $z$. 
The reason for this is that to lowest order (LO) in QCD, it turns out that $z$ is identical to a kinematic (scaling)
variable of the process. For instance, in $e^+e^-\to hX$ one has $z=2p_h\cdot q/q^2$ to LO, where $p_h$
is the momentum of the observed hadron and $q$ the momentum of the virtual photon that is produced by the 
$e^+e^-$ annihilation. NLO corrections dilute this direct ``local'' sensitivity only little. 
A drawback of $e^+e^-\to hX$ or $ep\to e h X$ is on the other hand that the gluon 
fragmentation function can be probed only indirectly by evolution or higher order corrections. 

Being universal objects, the same fragmentation functions are also relevant for describing hadron production in 
$pp$-scattering. So far, one has been using the process $pp\to hX$ as a further source of
information on the $D_c^h(z,\mu)$~\cite{dss07,dss14,Albino:2008fy}. Although this process does probe gluon fragmentation, 
its sensitivity to fragmentation functions is much less clear-cut than in the case of $e^+e^-\to hX$ or $ep\to e h X$. 
This is because for the single-inclusive process $pp\to hX$ the fragmentation functions arise in a more complex 
convolution with the partonic hard-scattering functions, which involves an integration over a typically rather wide range 
of $z$ already at LO. As a result, information on the $D_c^h(z,\mu)$ is smeared out and not readily available at a 
given fixed value of $z$.

The process $pp\rightarrow (\jet \,h)X$ allows to overcome this shortcoming. As it turns out, if one writes
its cross section differential in the variable $z_h$ introduced above, then to LO the hadron's
fragmentation function is to be evaluated at $z=z_h$. This means that by selecting $z_h$ 
one can ``dial'' the value at which the $D_c^h(z,\mu)$ are probed, similarly to what is available in 
$e^+e^-\to hX$ or $ep\to e h X$. Thanks to the fact that in $pp$ scattering different weights are given to 
the various fragmentation functions than compared to $e^+e^-\to hX$ and $ep\to e h X$, it is clear that 
$pp\rightarrow (\jet \,h)X$ has the potential to provide complementary new information on the 
$D_c^h(z,\mu)$, especially on gluon fragmentation. Data for $pp\rightarrow (\jet \,h)X$ should 
thus become valuable input to global QCD analyses of fragmentation functions. At the very least,
they should enable novel tests of the universality of fragmentation functions. We note that 
similar opportunities are expected to arise when the hadron is produced on the ``away-side''
of the jet, that is, basically back-to-back with the jet~\cite{deFlorian:2009fw}, although the 
kinematics is somewhat more elaborate in this case.

The production of specific hadrons inside jets  may also provide new insights into of the structure of 
jets and the hadronization mechanism. Varying $z_h$ and/or the hadron species, 
one can map out the abundances of specific hadrons in jets. Particle identification in jets becomes 
particularly interesting in a nuclear environment in $AA$ scattering, where distributions of hadrons 
may shed further light on the phenomenon of  ``jet quenching''.  Knowledge of fragmentation functions
in jet production and a good theoretical understanding of the process $pp\rightarrow (\jet \,h)X$ are
also crucial for studies of the Collins effect~\cite{Yuan:2008yv,Yuan:2008tv,D'Alesio:2010am}, 
an important probe of spin phenomena in hadronic scattering~\cite{Collins:1992kk}. 

In the present paper, we perform a new next-to-leading order (NLO) calculation of $pp\rightarrow (\jet \,h)X$.
In contrast to the previous calculation~\cite{Arleo:2013tya} which was entirely based on a numerical 
Monte-Carlo integration approach, we will derive analytical results for the relevant partonic cross sections. 
Apart from providing independent NLO predictions in a numerically very efficient way, this offers
several advantages. In the context of the analytical calculation, one can first of all 
explicitly check that the final-state collinear singularities have the structure required by the universality of 
fragmentation functions, meaning that the same fragmentation functions occur for $pp\rightarrow (\jet \,h)X$
as for usual single-inclusive processes such as $pp\rightarrow hX$. We note that to our knowledge
this has not yet been formally proven beyond NLO. Also, as we shall see, the NLO expressions show
logarithmic enhancements at high $z_h$, which recur with increasing power at every order in perturbation 
theory, eventually requiring resummation to all orders. Having explicit analytical results is a prerequisite
for such a resummation. In Ref.~\cite{Procura:2011aq}, considering the simpler case of $e^+e^-$-annihilation, 
such resummation calculations for large $z_h$ were presented. 

Technically, we will derive our results by assuming the jet to be relatively narrow, an approximation
known as ``Narrow Jet Approximation (NJA)''. This technique was
used previously for NLO calculations of single-inclusive jet production in hadronic scattering,
$pp\rightarrow \jet \,X$~\cite{Aversa:1990ww,Guillet:1990ez,Jager:2004jh,Mukherjee:2012uz,Kaufmann:2014nda}.
The main idea is to start from NLO ``inclusive-parton'' cross sections $d\hat{\sigma}_{ab}^c$ 
for the processes $ab\to cX$, which are relevant for the cross section for $pp\to hX$. They are
a priori not suitable for computing a jet cross section, which is evident from the fact that the
$d\hat{\sigma}_{ab}^c$ require collinear subtraction of final-state collinear singularities, whereas
a jet cross section is infrared-safe as far as the final state is concerned. Instead, it depends on the 
algorithm adopted to define the jet and thereby on a generic jet (size) parameter ${\cal R}$. As was shown
in Refs.~\cite{Aversa:1990ww,Guillet:1990ez,Jager:2004jh,Mukherjee:2012uz,Kaufmann:2014nda},
at NLO one may nonetheless go rather straightforwardly from the single-inclusive parton cross sections to the
jet ones, for any infrared-safe jet algorithm. The key is to properly account for the fact that at NLO two partons 
can fall into the same jet, so that the jet needs to be constructed from both. In fact, within the NJA, 
one can derive the translation between the $d\hat{\sigma}_{ab}^c$ and the partonic cross sections for
jet production analytically. We note that the NJA formally corresponds to the limit ${\cal R}\to 0$, but turns out to 
be accurate even at values ${\cal R}\sim 0.4-0.7$ relevant for experiment. In the NJA, the structure of the NLO 
jet cross section is of the form ${\cal A}\log(R)+{\cal B}$; corrections to this are of ${\cal O}({\cal R}^2)$ and are neglected. 
In this paper, we apply the NJA to the case of $pp\rightarrow (\jet \,h)X$, using it to derive the relevant
NLO partonic cross sections. In the course of the explicit NLO calculation, we find that the partonic cross 
sections for $pp\rightarrow \jet \,X$ and $pp\rightarrow (\jet \,h)X$ may be very compactly formulated 
in terms of the single-inclusive parton ones $d\hat{\sigma}_{ab}^c$, convoluted with appropriate perturbative 
``jet functions''. These functions are universal in the sense that they only depend on the type of the outgoing
partons that fragment and/or produce the jet, but not on the underlying partonic hard-scattering function.
On the basis of the jet functions, the NLO  partonic cross sections for $pp\rightarrow \jet \,X$ and 
$pp\rightarrow (\jet \,h)X$ take a very simple and systematic form. In fact, it turns out that for 
$pp\rightarrow (\jet \,h)X$ the jet functions have a ``two-tier'' form, with a first jet function describing
the formation of the jet and a second one the fragmentation of a parton inside the jet. We note that the 
concept of jet functions for formulating jet cross sections is not new but was introduced in the context of
soft-collinear effective theories 
(SCET)~\cite{Procura:2009vm,Jain:2011xz,Procura:2011aq,Bauer:2008jx,Ellis:2010rwa,Waalewijn:2012sv},
although applications to $pp\rightarrow (\jet \,h)X$ have to our knowledge not been given. 
Jet functions in a more general context of SCET or QCD resummation have been considered in
Refs.~\cite{Bauer:2003pi} and~\cite{Korchemsky:1992xv}, for example. We also note that
in Ref.~\cite{deFlorian:2009fw} the NLO corrections for the case of away-side jet-hadron correlations
were presented in the context of a Monte-Carlo integration code.

Our paper is organized as follows. In Sec.~\ref{NLOsec} we present our NLO calculation. In particular,
Sec.~\ref{hadinjet} contains our main new result, the formulation of $pp\rightarrow (\jet \,h)X$ in terms
of suitable jet functions. Section~\ref{pheno} presents phenomenological results for $pp\rightarrow (\jet \,h)X$
for LHC and RHIC. We finally conclude our work in Sec.~\ref{Concl}. The Appendices collect some
technical details of our calculations.

\section{Associated jet-plus-hadron production in the NJA \label{NLOsec}}

\subsection{Single-inclusive hadron production in hadronic collisions \label{had}}

Our formalism is best developed by first considering the process $H_1H_2\to h X$, where a hadron $h$
is observed at large transverse momentum $p_T$, but no requirement of a reconstructed hadronic jet is made. 
This is of course a standard reaction, for which the NLO corrections  have been known for a long 
time~\cite{Aversa:1988vb,Jager:2002xm}. 
The factorized cross section at given hadron $p_T$ and rapidity $\eta$ reads
\beeq\label{crsec1}
\frac{d\sigma^{H_1 H_2\to h X}}{dp_Td\eta}&=&\frac{2p_T}{S}
\sum_{abc} \int_{x_a^{\mathrm{min}}}^1 \frac{dx_a}{x_a} f_a^{H_1}(x_a,\mu_F)
\int_{x_b^{\mathrm{min}}}^1 \frac{dx_b}{x_b}  f_b^{H_2}(x_b,\mu_F)\nn\\[2mm]
&\times&\int_{z_c^{\mathrm{min}}}^1 \frac{dz_c}{z_c^2} \,
\frac{d\hat{\sigma}_{ab}^c(\hat{s},\hat{p}_T,\hat\eta,\mu_F,\mu_F',\mu_R)}{vdvdw}\,
D_c^h(z_c,\mu_F')\;,
\eeeq
with the usual parton distribution functions $f_a^H$, the fragmentation functions $D_c^h$,
and the hard-scattering cross sections $d\hat{\sigma}_{ab}^c$ for the partonic processes
$ab\to cX'$, $X'$ denoting an unobserved partonic final state. Defining
\beq\label{VW}
V \equiv 1 - \frac{p_T}{\sqrt{S}}\, e^{-\eta} \,,\quad W \equiv \frac{p_T^2}{SV(1-V)}\,,
\eeq
with $\sqrt{S}$ the hadronic c.m.s. energy, we have
\beq\label{xaxbzc}
x_a^{\mathrm{min}}\,=\,W\,,\quad x_b^{\mathrm{min}}\,=\,\frac{1-V}{1-VW/x_a}\,,\quad
z_c^{\mathrm{min}}\,=\,\frac{1-V}{x_b}+\frac{VW}{x_a}\,.
\eeq
The $d\hat{\sigma}_{ab}^c$ are functions of the partonic
c.m.s. energy $\hat{s}=x_ax_bS$, the partonic transverse momentum 
$\hat{p}_T=p_T/z_c$ and the partonic rapidity $\hat\eta=\eta-\frac{1}{2}\log(x_a/x_b)$.
Since only $\hat{p}_T$ depends on $z_c$, the last integral in Eq.~(\ref{crsec1}) 
takes the form of a convolution. The variables $v$ and $w$ in~(\ref{crsec1}) are the partonic 
counterparts of $V$ and $W$: 
\beq\label{svsw}
v \equiv 1-\frac{\hat{p}_T {\mathrm{e}}^{-\hat\eta}}{\sqrt{\hat{s}}}\, , \quad
w \equiv \frac{\hat{p}_T^2}{\hat{s}v(1-v)}\;.
\eeq
One customarily expresses $\hat{p}_T$ and $\hat\eta$ by $v$ and $w$:
\beq
\hat{p}_T^2=\hat{s}vw(1-v)\;\;,\;\;\;\;\hat\eta=\frac{1}{2}\log\left(\frac{vw}{1-v}\right)\,.
\eeq
Finally, the various functions in Eq.~(\ref{crsec1}) are tied together by their dependence on 
the initial- and final-state factorization scales, $\mu_F$ and $\mu_F'$, respectively, 
and the renormalization scale $\mu_R$. 

The partonic hard-scattering cross sections may be evaluated in QCD perturbation theory. 
We write the perturbative expansion to NLO as 
\beq\label{sigpert}
\frac{d\hat{\sigma}_{ab}^c}{dvdw}\,=\,
\frac{d\hat{\sigma}_{ab}^{c,(0)}}{dv} \,\delta(1-w)+
\frac{\alpha_{s}(\mu_R)}{2\pi} \frac{d\hat{\sigma}^{c,(1)}_{ab}}{dvdw}+{\cal O}(\alpha_s^2(\mu_R))\,,
\eeq
where we have used that $w=1$ for leading-order (LO) kinematics (since the unobserved partonic 
final state $X'$ consists of a single parton), equivalent to $2\hat{p}_T\cosh(\hat{\eta})/\sqrt{\hat{s}}=1$. 
The NLO terms $d\hat{\sigma}^{c,(1)}_{ab}$ have been presented in Refs.~\cite{Aversa:1988vb,Jager:2002xm}.

\subsection{Translation to single-inclusive jet cross section via jet functions}

As shown in Refs.~\cite{Aversa:1990ww,Guillet:1990ez,Jager:2004jh,Mukherjee:2012uz}, 
one can transform the cross section for single-inclusive hadron production to a 
single-inclusive jet one. References~\cite{Aversa:1990ww,Guillet:1990ez,Jager:2004jh,Mukherjee:2012uz}
explicitly constructed this translation at NLO. We may write the jet cross section as
\beq\label{crsec2}
\frac{d\sigma^{H_1 H_2\to {\mathrm{jet}} X}}{dp_T^{\mathrm{jet}} d\eta^{\mathrm{jet}}} \,=\,
\frac{2p_T^{\mathrm{jet}}}{S}
\sum_{ab} \int_{x_a^{\mathrm{min}}}^1 \frac{dx_a}{x_a} f_a^{H_1}(x_a,\mu_F)
\int_{x_b^{\mathrm{min}}}^1 \frac{dx_b}{x_b}  f_b^{H_2}(x_b,\mu_F)
\frac{d\hat{\sigma}_{ab}^{\mathrm{jet,algo}}(\hat{s},p_T^{\mathrm{jet}},\hat\eta,\mu_F,\mu_R,{\cal R})}{vdvdw}\;,
\eeq
where $p_T^{\mathrm{jet}}$ and $\eta^{\mathrm{jet}}$ are the jet's transverse momentum
and rapidity, and where ${\cal R}$ denotes a parameter specifying the jet algorithm. For the
jet cross section we still have 
\beq\label{xaxb}
x_a^{\mathrm{min}}\,=\,W\,,\quad x_b^{\mathrm{min}}\,=\,\frac{1-V}{1-VW/x_a}\,,
\eeq
as in~(\ref{xaxbzc}), but with $V$ and $W$ now defined by
\beq\label{VW1}
V \equiv 1 - \frac{p_T^{\mathrm{jet}}}{\sqrt{S}}\, e^{-\eta^{\mathrm{jet}}} \,,\quad 
W \equiv \frac{(p_T^{\mathrm{jet}})^2}{SV(1-V)}\,.
\eeq
Likewise, $v$ and $w$ are as in~(\ref{svsw}) but with $\hat{p}_T\to p_T^{\mathrm{jet}}$.
Furthermore, in analogy with the inclusive-hadron case, $\hat\eta=\eta^{\mathrm{jet}}-
\frac{1}{2}\log(x_a/x_b)$. We note that the partonic cross sections $d\hat{\sigma}_{ab}^{\mathrm{jet,algo}}$
relevant for jet production depend on the algorithm used to define the jet. They do not carry any dependence 
on a final-state factorization scale. 

In order to go from the inclusive-parton cross sections $d\hat{\sigma}_{ab}^c$ to the jet
ones $d\hat{\sigma}_{ab}^{\mathrm{jet}}$, the idea is to apply proper correction terms to the 
former. The $d\hat{\sigma}_{ab}^c$ 
have been integrated over the full phase space of all final-state partons other than $c$. Therefore, 
they contain contributions where a second parton in the final state is so close to parton $c$ that the two should
jointly form the jet for a given jet definition. One can correct for this by subtracting 
such contributions from $d\hat{\sigma}_{ab}^c$ and adding a piece where they actually do form the jet together. At 
NLO, where there can be three partons $c,d,e$ in the final state, one has after suitable summation over all
possible configurations:
\begin{eqnarray}\label{deco}
d \hat{\sigma}_{ab}^{\mathrm{jet}} &=&
[d\hat{\sigma}_{ab}^c -d \hat{\sigma}_{ab}^{c(d)}-
d \hat{\sigma}_{ab}^{c(e)}]+ [d\hat{\sigma}_{ab}^d -d\hat{\sigma}_{ab}^{d(c)}-
d \hat{\sigma}_{ab}^{d(e)}]+
[d\hat{\sigma}_{ab}^e -d \hat{\sigma}_{ab}^{e(c)}-
d \hat{\sigma}_{ab}^{e(d)}]\nonumber\\[2mm] &+&
d \hat{\sigma}_{ab}^{cd} + d \hat{\sigma}_{ab}^{ce}+
d \hat{\sigma}_{ab}^{de} \, .
\label{jetform}
\end{eqnarray}
Here $d\hat{\sigma}_{ab}^{j(k)}$ is the cross section where parton $j$ produces the jet, but parton 
$k$ is so close that it should be part of the jet, and $d \hat{\sigma}_{ab}^{jk}$ is the cross section 
when both partons $j$ and $k$ jointly form the jet. The decomposition~(\ref{deco}) is completely 
general to NLO. It may be applied for any jet algorithm, as long as the algorithm is infrared-safe. 
As mentioned before, a property of the $d\hat{\sigma}_{ab}^{\mathrm{jet}}$
is that all dependence on the final-state factorization scale $\mu_F'$, which was initially
present in the $d\hat{\sigma}_{ab}^j$, must cancel. This cancellation comes about in~(\ref{deco})
because the $d \hat{\sigma}_{ab}^{jk}$ possess final-state collinear singularities that require factorization.
This introduces dependence on $\mu_F'$ in exactly the right way as to compensate the
$\mu_F'$-dependence of the $d\hat{\sigma}_{ab}^j$. 

In the NJA, the correction terms $d\hat{\sigma}_{ab}^{j(k)}$ and $d \hat{\sigma}_{ab}^{jk}$
may be computed analytically. At NLO, they both receive contributions from real-emission 
$2\to 3$ diagrams only. For the NJA one assumes that the observed jet is rather collimated. 
This in essence allows to treat the two outgoing partons $j$ and $k$ as collinear.
The relevant calculations for the standard cone\footnote{Here we have in mind primarily
the ``Seedless Infrared Safe Cone'' (SISCone) algorithm introduced in Ref.~\cite{Salam:2007xv} 
which represents the only cone-based jet definition known to be strictly infrared-safe. However,
for single-inclusive jet cross sections, the lack of infrared-safety of other cone-type algorithms 
occurs first at next-to-next-to-leading order in perturbation theory and hence is not an issue here.} 
and (anti-)$k_t$~\cite{Ellis:1993tq,Catani:1993hr,Cacciari:2008gp}  algorithms were carried out in 
Refs.~\cite{Jager:2004jh,Mukherjee:2012uz}, while Ref.~\cite{Kaufmann:2014nda} addressed 
the case of the ``$J_{E_T}$'' algorithm proposed in~\cite{Georgi:2014zwa,Bai:2014qca}.  
We note that we always define the four-momentum of the jet as the sum of four-momenta of the 
partons that form the jet. This so-called ``$E$ recombination scheme''~\cite{escheme} 
is the most popular choice nowadays.

By close inspection of~(\ref{deco}), we have found that in the NJA the jet cross section may be cast
into a form that makes use of the {\it single-inclusive parton} production cross sections 
$d\hat{\sigma}_{ab}^c$:
\beeq\label{crsec3}
\frac{d\sigma^{H_1 H_2\to {\mathrm{jet}} X}}{dp_T^{\mathrm{jet}} d\eta^{\mathrm{jet}}} 
&=&\frac{2p_T^{\mathrm{jet}}}{S}
\sum_{abc} \int_{x_a^{\mathrm{min}}}^1 \frac{dx_a}{x_a} f_a^{H_1}(x_a,\mu_F)
\int_{x_b^{\mathrm{min}}}^1 \frac{dx_b}{x_b}  f_b^{H_2}(x_b,\mu_F)\nn\\[2mm]
&\times&\int_{z_c^{\mathrm{min}}}^1 \frac{dz_c}{z_c^2} \,
\frac{d\hat{\sigma}_{ab}^c(\hat{s},\hat{p}_T,\hat\eta,\mu_F,\mu_F',\mu_R)}{vdvdw}\,
{\cal J}_c\left(z_c,\frac{{\cal R}\,p_T^{\mathrm{jet}}}{\mu_F'},\mu_R\right)\;,
\eeeq
with {\it inclusive jet functions} ${\cal J}_q$ and ${\cal J}_g$. We have $\hat{p}_T=
p_T^{\mathrm{jet}}/z_c$, and $x_a^{\mathrm{min}},x_b^{\mathrm{min}},z_c^{\mathrm{min}}$ 
and~$v,w$ are now as in~(\ref{xaxbzc}) and~(\ref{svsw}), respectively. Equation~(\ref{crsec3}) thus 
states that one can go directly from the cross section for single-hadron production to that
for jet production by replacing the fragmentation functions $D_c^h$ in~(\ref{crsec1}) by
the jet functions ${\cal J}_c$. The latter are such that any dependence on $\mu_F'$ 
disappears from the cross section. They depend on the jet algorithm and hence on a 
jet parameter ${\cal R}$. For the cone and (anti-)$k_t$ algorithms ${\cal R}$ 
is just given by the usual jet size parameter $R$ introduced for these algorithms, while
for the jet algorithm of~\cite{Georgi:2014zwa,Bai:2014qca} 
we have ${\cal R}=1/\sqrt{\beta z_c}$ with $\beta$ the ``maximization''
parameter defined for this algorithm. In the NJA we generally assume ${\cal R}\ll 1$ and 
neglect $\mathcal{O}({\cal R}^2)$ contributions.  The jet functions then read explicitly
\beeq\label{Jinc}
{\cal J}_q\left(z,\lambda\equiv \frac{{\cal R}\,p_T^{\mathrm{jet}}}{\mu_F'},\mu_R\right)&=&
\delta(1-z)-\frac{\alpha_s(\mu_R)}{2\pi}\left[2 C_F (1+z^2)\left(\frac{\log(1-z)}{1-z}\right)_+
+P_{qq}(z)\log\left(\lambda^2\right)\right.\nn\\[2mm]
&&\hspace*{3.6cm}+\,\delta(1-z)I_q^{\mathrm{algo}}+C_F(1-z)\bigg]\nn\\[2mm]
&&\hspace*{1.6cm}-\,\frac{\alpha_s(\mu_R)}{2\pi}\Big[
P_{gq}(z)\log\left( \lambda^2(1-z)^2\right)+ C_F\,z\Big]\,,\nn\\[3mm]
{\cal J}_g\left(z,\lambda\equiv \frac{{\cal R}\,p_T^{\mathrm{jet}}}{\mu_F'},\mu_R\right)&=&
\delta(1-z)-\frac{\alpha_s(\mu_R)}{2\pi}\,\left[\frac{4C_A(1-z+z^2)^2}{z}
\left(\frac{\log(1-z)}{1-z}\right)_+\right.\nn\\[2mm]
&&\hspace*{3.7cm}+\,P_{gg}(z)\log\left(\lambda^2\right)+\delta(1-z)I_g^{\mathrm{algo}}\bigg]\nn\\[2mm]
&&\hspace*{1.6cm}-\,\frac{\alpha_s(\mu_R)}{2\pi}\,2n_f\Big[
P_{qg}(z)\log\left( \lambda^2(1-z)^2\right)+ z(1-z)\Big]\,,
\eeeq
where $C_F=4/3$, $C_A=3$ and $n_f$ is the number of active flavors, 
and where the LO splitting functions $P_{ij}(z)$ as well as 
the ``plus''-distribution are defined in Appendix~\ref{appendix:jetfuncs}. The dependence
on the jet algorithm is reflected in the terms $I_q^{\mathrm{algo}}$ and $I_g^{\mathrm{algo}}$, 
which are just numbers that we also collect in Appendix~\ref{appendix:jetfuncs}.

Equation~(\ref{crsec3}) evidently exhibits a factorized structure in the final state for the jet
cross section in the NJA. Its physical interpretation is essentially that the hard scattering produces
a parton $c$ that ``fragments'' into the observed jet via the jet function ${\cal J}_c$,
the jet carrying the fraction $z_c$ of the produced parton's momentum. At NLO, the 
factorization is in fact rather trivial. To get a clear sense of it, it is instructive to see how one
recovers~(\ref{crsec2}),(\ref{deco}) from~(\ref{crsec3}). To this end, we combine~(\ref{sigpert}) and~(\ref{Jinc})
and expand to first order in the strong coupling. The products of the $d\hat{\sigma}_{ab}^c$ with 
the LO $\delta(1-z)$ terms in ${\cal J}_c$ just reproduce the single-inclusive
parton cross sections at $\hat{p}_T=p_T^{\mathrm{jet}}$, i.e. the terms $d\hat{\sigma}_{ab}^c,
d\hat{\sigma}_{ab}^d,d\hat{\sigma}_{ab}^e$ in~(\ref{deco}). The only other terms surviving in the 
expansion to ${\cal O}(\alpha_s)$ are the products of the LO terms $\delta(1-w)\,
d\hat{\sigma}_{ab}^{c,(0)}/dv$ of~(\ref{sigpert}) with the ${\cal O}(\alpha_s)$ terms in the jet functions. 
These precisely give the remaining contributions $d \hat{\sigma}_{ab}^{cd}-d \hat{\sigma}_{ab}^{c(d)}-
d \hat{\sigma}_{ab}^{d(c)}$ (plus the other combinations) in~(\ref{deco}). Because of the convolution
in $z_c$ in~(\ref{crsec3}), the $\delta(1-w)$-function in the Born cross section actually fixes
$z_c$ to the value $z_c=2p_T^{\mathrm{jet}}\cosh(\hat\eta)/\sqrt{\hat{s}}$. Based on our NLO calculation,
we evidently cannot prove the factorization shown in~(\ref{crsec3}) to beyond this order. We note, however,
that similar factorization formulas have been derived using Soft Collinear Effective Theory (SCET) 
techniques~\cite{Ellis:2010rwa,Bauer:2008jx,Waalewijn:2012sv},
for the case of jet observables in $e^+e^-$ annihilation. In particular, functions closely related to our
inclusive jet functions ${\cal J}_{q,g}$ may be found in Ref.~\cite{Ellis:2010rwa}, where they are termed 
``unmeasured'' quark (or gluon) jet functions. We shall return to comparisons with SCET results
below. 

\subsection{Hadrons produced inside jets \label{hadinjet}}

We are now ready to tackle the case that we are really interested in, $H_1 H_2\rightarrow (\jet \,h)X$ 
where the hadron is observed inside a reconstructed jet and is part of the jet. 
Our strategy for performing an analytical NLO
calculation will be to use the NJA and the same considerations as those that gave rise to Eq.~(\ref{deco}).
Subsequently, we will again phrase our results in a simple and rather general way in terms of suitable jet functions. 

The cross section we are interested in is specified by the jet's transverse momentum $p_T^{\mathrm{jet}}$
and rapidity $\eta^{\mathrm{jet}}$, and by the variable 
\beq\label{zrel}
z_h\,\equiv\,\frac{p_T}{p_T^{\mathrm{jet}}}\,,
\eeq
where as in section~\ref{had} $p_T$ refers to the transverse momentum of the produced hadron. 
As we are working in the NJA, we consider collinear fragmentation of the hadron inside the jet. Thus, 
the observed hadron and the jet have the same rapidities, $\eta = \eta^\jet$, since differences in 
rapidity are $\mathcal{O}({\cal R}^2)$ effects and hence suppressed in the NJA.  

The factorized jet-plus-hadron cross section is written as 
\beeq\label{Xsection}
\frac{d\sigma^{H_1H_2\rightarrow (\jet\,h)X}}{dp_T^{\jet} d\eta^\jet d z_h} &= &
\frac{2p_T^{\jet}}{S} \sum_{a,b,c} \int_{x_a^{\mathrm{min}}}^1 \frac{dx_a}{x_a} f_a^{H_1}(x_a,\mu_F)
\int_{x_b^{\mathrm{min}}}^1 \frac{dx_b}{x_b}  f_b^{H_2}(x_b,\mu_F)\nn\\[2mm]
&\times&\int_{z_h}^1 \frac{dz_p}{z_p}\,\frac{d\hat{\sigma}_{ab}^{\mathrm{(jet}\,c)}
(\hat{s},p_T^{\jet},\hat\eta,\mu_F,\mu_F',\mu_R,{\cal R},z_p)}{vdvdwdz_p}\,
D_c^h\left(\frac{z_h}{z_p},\mu_F^\prime\right)\,,
\eeeq
where $x_a^{\mathrm{min}}$, $x_b^{\mathrm{min}}$, and $\hat\eta=\eta^{\mathrm{jet}}-\frac{1}{2}\log(x_a/x_b)$
are as for the single-inclusive jet cross section, and where $z_p$ is the partonic analog of $z_h$.
In other words, the $d\hat{\sigma}_{ab}^{\mathrm{(jet}\,c)}$ are the partonic cross sections for producing 
a final-state jet (subject to a specified jet algorithm), inside of which there is a parton $c$ with transverse
momentum $p_T^c=z_p p_T^{\jet}$ that fragments into the observed hadron. The argument of the corresponding
fragmentation functions is fixed by $p_T=z p_T^c$ and hence, using~(\ref{zrel}), is given by $z=z_h/z_p$. 
Thus the new partonic cross sections are in convolution with the fragmentation functions. Note that 
all other variables $V,W$ and $v,w$ have the same definitions as in the single-inclusive jet case; 
see Eq.~(\ref{VW1}). 

At lowest order, there is only one parton forming the jet, and this parton also is the one that
fragments into the observed hadron, implying $z_p=1$. The partonic cross sections hence have the 
perturbative expansions 
\beq
\frac{d\hat{\sigma}_{ab}^{\mathrm{(jet}\,c)}}{dvdwdz_p}\,=\,
\frac{d\hat{\sigma}_{ab}^{c,(0)}}{dv} \,\delta(1-w)\,\delta(1-z_p) + 
\frac{\alpha_s(\mu_R)}{\pi}\,\frac{d\hat{\sigma}_{ab}^{\mathrm{(jet}\,c),(1)}}{dvdwdz_p}  
+ \mathcal{O}(\alpha_s^2(\mu_R))\label{partonic_series}\,,
\eeq
with the same Born terms $d\hat{\sigma}_{ab}^{c,(0)}/dv$ as in~(\ref{sigpert}). 

In order to derive the NLO partonic cross sections $d\hat{\sigma}_{ab}^{\mathrm{(jet}\,c),(1)}$, 
we revisit Eq.~(\ref{deco}). Since we now ``observe'' a parton $c$ in the final state (the fragmenting one), 
we must not sum over all possible final states, but rather consider only the contributions that contain parton $c$:
\beq
d\hat{\sigma}_{ab}^c -d \hat{\sigma}_{ab}^{c(d)}-
d \hat{\sigma}_{ab}^{c(e)}+d \hat{\sigma}_{ab}^{cd} + d \hat{\sigma}_{ab}^{ce} \,.
\label{sigma:organization}
\eeq
However, for each term we now need to derive its proper dependence on $z_p$ before
combining all terms. 
For the terms $d\hat{\sigma}_{ab}^c$ and $d \hat{\sigma}_{ab}^{c(d)},d \hat{\sigma}_{ab}^{c(e)}$
this is trivial since for all of these terms parton $c$ alone produces the jet and also 
is the parton that fragments. As a result, all these terms simply acquire a factor $\delta(1-z_p)$. 
This becomes different for the pieces $d\hat{\sigma}_{ab}^{cd} ,d \hat{\sigma}_{ab}^{ce}$.
Following~\cite{Jager:2004jh,Mukherjee:2012uz,Kaufmann:2014nda}, in the NJA we may write 
the NLO contribution to any $d\hat{\sigma}_{ab}^{cd}$ as 
\beq
\label{eq:scaps5}
\frac{d\hat{\sigma}_{ab}^{cd,(1)}}{dvdw}\,=\,\frac{\alpha_s}{\pi}\,{\cal N}_{ab\to K}
(v,w,\varepsilon)\,\delta(1-w) \int_0^1 dz_p\, z_p^{-\varepsilon} (1-z_p)^{-\varepsilon}\tilde{P}_{cK}^{<}(z_p)
\,\int_0^{m^2_{{\mathrm{max,algo}}}} \frac{dm_\jet^2}{m_\jet^2} \,m_\jet^{-2 \varepsilon},
\eeq
where we have used dimensional regularization with $D=4-2\varepsilon$ space-time dimensions.
Equation~(\ref{eq:scaps5}) is derived from the fact that the leading contributions in the NJA come
from a parton $K$ splitting into partons $c$ and $d$ ``almost'' collinearly in the final state. We therefore have 
an underlying Born process $ab\to KX$ (with some unobserved recoil final state $X$), whose $D$-dimensional 
cross section is contained in the ``normalization factor'' ${\cal N}_{ab\to K}$, along with some trivial factors. 
The integrand then contains the $D$-dimensional LO splitting functions $\tilde{P}_{cK}^<(z)$,
where the superscript ``$<$'' indicates that the splitting function is strictly at $z<1$, that is, without
its $\delta(1-z)$ contribution that is present when $c=K$. The functions are defined in Eq.~(\ref{PT1})
in Appendix~\ref{appendix:jetfuncs}.
The argument of the splitting function is the fraction of the intermediate particle's momentum 
(equal to the jet momentum) transferred in the splitting. In the NJA it therefore coincides with our partonic
variable $z_p$. In the second integral in~(\ref{eq:scaps5}) $m_\jet$ is the invariant mass of the jet. 
The explicit factor $m_\jet^2$ in the denominator represents the propagator of the splitting parton $K$. 
The integral over the jet mass runs between zero and an upper limit $m_{{\mathrm{max,algo}}}$, which 
in the NJA is formally taken to be relatively small. As indicated, $m_{{\mathrm{max,algo}}}$ depends on the 
algorithm chosen to define the jet. We have~\cite{Jager:2004jh,Mukherjee:2012uz,Kaufmann:2014nda}
\beq
m_{{\mathrm{max,algo}}}^2 \,=\,\left\{\begin{array}{cc}
(p_T^{\mathrm{jet}}R)^2\,\min\left(\frac{z_p}{1-z_p},\frac{1-z_p}{z_p}\right)& {\mathrm{cone\;algorithm}}\,,\\[2mm]
(p_T^{\mathrm{jet}}R)^2\,z_p(1-z_p)& {\mathrm{(anti-)}}k_t {\mathrm{\;algorithm}}\,,\\[2mm]
\frac{(p_T^{\mathrm{jet}})^2}{\beta}\,\min(z_p,1-z_p)& J_{E_T} {\mathrm{\;algorithm}}\,.
\end{array}\right.
\eeq
To make the cross section $d\hat{\sigma}_{ab}^{cd,(1)}$ differential in $z_p$ we now just need to drop
the integration over $z_p$ in~(\ref{eq:scaps5}). We next expand the resulting expression in $\varepsilon$.
The $m_\jet^2$ integration produces a collinear singularity in $1/\varepsilon$. It also contributes
a factor $(1-z_p)^{-\varepsilon}$ at large $z_p$ which may be combined with the explicit factor 
$(1-z_p)^{- \varepsilon}$  in~(\ref{eq:scaps5}). In the presence of a diagonal splitting function in the 
integrand we hence arrive at a term $(1-z_p)^{-1-2\varepsilon}$, which may be expanded
in $\varepsilon$ to give a further pole in $1/\varepsilon$ and ``plus''-distributions in $1-z_p$. 
The double poles $1/\varepsilon^2$ arising in this way cancel against double poles in
$d \hat{\sigma}_{ab}^{c(d)},d \hat{\sigma}_{ab}^{c(e)}$. The remaining single poles 
are removed by collinear factorization into the fragmentation function for parton $c$. For
non-diagonal splitting functions there are only single poles which are directly subtracted by factorization. 
We note that the original $d\hat{\sigma}_{ab}^{cd,(1)}$ is in fact needed both for the cross section with parton 
$c$ fragmenting and also for the one where $d$ fragments. This is reflected in the fact that the $z_p$-integral 
in~(\ref{eq:scaps5}) runs from $0$ to $1$, while for $d\hat{\sigma}_{ab}^{cd,(1)}$ 
the limit $z_p\to 0$ is never reached as long as $z_h>0$. For parton $d$ fragmenting, however, 
we need to use $d\hat{\sigma}_{ab}^{dc,(1)}$ which differs from $d\hat{\sigma}_{ab}^{cd,(1)}$ only
by a change of the splitting function. In case of a quark splitting into a quark and a gluon, this change
is from $\tilde{P}^<_{qq}(z)$ for an observed quark to $\tilde{P}^<_{gq}(z)$ for an observed gluon.
Because of $\tilde{P}^<_{gq}(z)=\tilde{P}^<_{qq}(1-z)$ one precisely recovers the old expression
for the inclusive-jet cross sections when all final states are summed over. Likewise, if a gluon
splits into a $q\bar{q}$ or $gg$, the relevant splitting functions $\tilde{P}^<_{qg}(z)$, $\tilde{P}^<_{gg}(z)$
are by themselves symmetric under $z\leftrightarrow 1-z$. 

From this discussion, and combining with Eqs.~(\ref{partonic_series}),(\ref{sigma:organization}),
we obtain to NLO in the NJA:
\beq
\frac{d\hat{\sigma}_{ab}^{\mathrm{(jet}\,c)}}{dvdwdz_p}\,=\,
\left[\frac{d\hat{\sigma}_{ab}^c}{dvdw}-\frac{d\hat{\sigma}_{ab}^{c(d)}}{dvdw}
-\frac{d\hat{\sigma}_{ab}^{c(e)}}{dvdw}\right]\delta(1-z_p) + 
\frac{d\hat{\sigma}_{ab}^{cd}}{dvdwdz_p}+\frac{d\hat{\sigma}_{ab}^{ce}}{dvdwdz_p}\,.
\eeq
Computing and inserting all ingredients of this expression, we find that the cross section may
be cast into a form that again makes use of the {\it single-inclusive parton} production cross sections 
$d\hat{\sigma}_{ab}^c$, similar to the case of inclusive-jet production in Eq.~(\ref{crsec3}):
\beeq\label{Xsection1}
\frac{d\sigma^{H_1H_2\rightarrow (\jet\,h)X}}{dp_T^{\jet} d\eta^\jet d z_h} &= &
\frac{2p_T^{\jet}}{S} \sum_{a,b,c} \int_{x_a^{\mathrm{min}}}^1 \frac{dx_a}{x_a} f_a^{H_1}(x_a,\mu_F)
\int_{x_b^{\mathrm{min}}}^1 \frac{dx_b}{x_b}  f_b^{H_2}(x_b,\mu_F)\nn\\[2mm]
&&\hspace*{-3.3cm}\times\;\int_{z_c^{\mathrm{min}}}^1 \frac{dz_c}{z_c^2} \,
\frac{d\hat{\sigma}_{ab}^c(\hat{s},\hat{p}_T,\hat\eta,\mu_F,\mu_F',\mu_R)}{vdvdw}\,
\sum_{c'}\int_{z_h}^1 \frac{dz_p}{z_p}\,
{\cal K}_{c\to c'}\left(z_c,z_p;\frac{{\cal R}\,p_T^{\mathrm{jet}}}{\mu_F'},
\frac{{\cal R}\,p_T^{\mathrm{jet}}}{\mu_F''},\mu_R\right)
D_{c'}^h\left(\frac{z_h}{z_p},\mu_F''\right)\,,\nn\\
\eeeq
where $x_a^{\mathrm{min}},x_b^{\mathrm{min}}$ and $z_c^{\mathrm{min}}$ are as given in~(\ref{xaxbzc}),
with $V$ and $W$ defined in terms of jet transverse momentum and rapidity. Furthermore, as in~(\ref{crsec3})
we have $\hat{p}_T=p_T^{\mathrm{jet}}/z_c$. The (jet algorithm dependent) functions ${\cal K}_{c\to c'}$
are new ``semi-inclusive'' jet functions that describe the production of a fragmenting parton $c'$ inside a jet
that results from a parton $c$ produced in the hard scattering. For the ``transition'' $q\to q$ we find
\beeq \label{Kqq}
\mathcal{K}_{q\rightarrow q}\left(z,z_p,\lambda=\frac{{\cal R}\,p_T^{\mathrm{jet}}}{\mu_F'},
\kappa=\frac{{\cal R}\,p_T^{\mathrm{jet}}}{\mu_F''},\mu_R\right)&=&\delta(1-z)\delta(1-z_p)+
\frac{\alpha_s(\mu_R)}{2\pi}\bigg[\nonumber\\[2mm]
&&\hspace*{-7cm}-\,\delta(1-z_p)\left\{2 C_F (1+z^2)\left(\frac{\log(1-z)}{1-z}\right)_+
+\,P_{qq}(z)\log\left(\lambda^2\right)+C_F(1-z)\right\}\nonumber\\[2mm]
&&\hspace*{-7cm}+\;\delta(1-z) \left.\left\{ 2 C_F (1+z_p^2)\left(\frac{\log(1-z_p)}{1-z_p}\right)_+
+\,P_{qq}(z_p)\log\left(\kappa^2\right)+C_F(1-z_p) 
+\mathcal{I}_{qq}^\text{algo}(z_p) \right\}\right]\,,\nn\\
\eeeq
where $\mathcal{I}_{qq}^\text{algo}(z_p)$ is a function that depends on the jet algorithm. 
Since we will write our new jet functions in a more compact form below, we do not present
the other functions $\mathcal{K}_{c\rightarrow c'}$ 
here but collect them in Appendix~\ref{appendix:jetfuncs1}, along with the
$\mathcal{I}_{c'c}^\text{algo}(z_p)$. 

As indicated, $\mathcal{K}_{q\rightarrow q}$ carries dependence on two 
(final-state) factorization scales, $\mu_F'$ and $\mu_F''$. The former is the same as we 
encountered in the case of single-inclusive jets in Eqs.~(\ref{crsec3}),(\ref{Jinc}). It was
originally introduced in the collinear factorization for the single-inclusive parton cross
sections, but now has to cancel exactly between the $d\hat{\sigma}_{ab}^c$ and the 
$\mathcal{K}_{c\rightarrow c'}$. As in the case of single-inclusive jets,
the cancelation of dependence on $\mu_F'$ is just a result of the fact that we foremost
define our observable by requiring a jet in the final state. In this sense, $\mu_F'$ is 
simply an artifact of the way we organize the calculation and is not actually present in the
final answer. The scale $\mu_F''$, on the other hand, arises because we now also require a
hadron in the final state. Technically it arises when we subtract collinear singularities from 
the $d\hat{\sigma}_{ab}^{cd,(1)}$. The logarithms in $\mu_F''$ are thus just the standard
scale logarithms that compensate the evolution of the fragmentation functions at this order.
We also note that there are two sum rules that connect the inclusive and the semi-inclusive 
jet functions~\cite{Procura:2009vm,Jain:2011xz}:
\begin{align}
\int_0^1 dz_p\; z_p \left[\mathcal{K}_{q\to q}(z,z_p;\lambda,\kappa,\mu_R) + 
\mathcal{K}_{q\to g}(z,z_p;\lambda,\kappa,\mu_R)\right] &= 
\mathcal{J}_q(z,\lambda,\mu_R)\,,\nonumber\\[2mm]
\int_0^1 dz_p\; z_p \left[\mathcal{K}_{g\to g}(z,z_p;\lambda,\kappa,\mu_R) + 
\mathcal{K}_{g\to q}(z,z_p;\lambda,\kappa,\mu_R)\right] &= 
\mathcal{J}_g(z,\lambda,\mu_R)\,.
\end{align}
Both are fulfilled by our expressions. Furthermore, $\int_0^1 dz_p\,\mathcal{K}_{q\to q}
(z,z_p;\lambda,\kappa,\mu_R)$ reproduces the quark splitting contributions to $\mathcal{J}_q$,
i.e. the first two lines in Eq.~(\ref{Jinc}). 

\begin{figure}[t]
	\centering
	\includegraphics[width=0.7\textwidth]{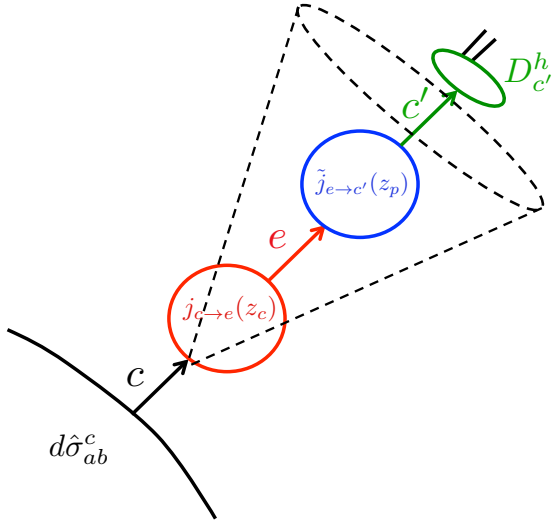}
	\vspace*{-1.5cm}
	\caption{{\sl Sketch of the production of an observed hadron inside a jet, described in terms of the 
	jet functions $j_{c\to e}$ and $\tilde{j}_{e\to c'}$ (see text).}}
	\label{JetFuncsPic}
\end{figure}

We may actually go one step further and decompose the functions 
$\mathcal{K}_{c\rightarrow c'}$ into products of jet functions that 
separate the dependence on $z$ and $z_p$. We define two sets of functions:
\begin{align}
j_{q\to q}\left(z,\lambda,\mu_R\right) &\equiv \delta(1-z)-\frac{\alpha_s(\mu_R)}{2\pi}\left[
2 C_F (1+z^2)\left(\frac{\log(1-z)}{1-z}\right)_++P_{qq}(z)\log\left(\lambda^2\right)\right.\nonumber\\[2mm]
& \hspace*{4.1cm}+\,\delta(1-z)I_q^{\mathrm{algo}}+C_F(1-z)\bigg]\,,\nonumber\\[2mm]
j_{q\to g}\left(z,\lambda,\mu_R\right) &\equiv-\frac{\alpha_s(\mu_R)}{2\pi}
\left[P_{gq}(z)\log\left( \lambda^2(1-z)^2\right)+ 
C_Fz\right]\,,\nonumber\\[2mm]
j_{g\to g}\left(z,\lambda,\mu_R\right) &\equiv\delta(1-z)-\frac{\alpha_s(\mu_R)}{2\pi}
\left[\frac{4C_A(1-z+z^2)^2}{z}
\left(\frac{\log(1-z)}{1-z}\right)_++P_{gg}(z)\log\left(\lambda^2\right)\right.\nn\\[2mm]
& \hspace*{4.1cm}+\,\delta(1-z)I_g^{\mathrm{algo}}
\bigg]\,,\nonumber\\[2mm]
j_{g\to q}\left(z,\lambda,\mu_R\right) &\equiv-\frac{\alpha_s(\mu_R)}{2\pi}\left[
P_{qg}(z)\log\left( \lambda^2(1-z)^2\right)+ z(1-z)\right]\,,
\end{align} 
(where as before $\lambda={\cal R}\,p_T^{\mathrm{jet}}/\mu_F'$), and
\begin{align}
\tilde{j}_{q\to q}\left(z_p,\kappa,\mu_R\right)&\equiv
\delta(1-z_p)+\frac{\alpha_s(\mu_R)}{2\pi}\left[2C_F (1+z_p^2)\left(\frac{\log(1-z_p)}{1-z_p}\right)_+
+\,P_{qq}(z_p)\log\left(\kappa^2\right)\right.\nn\\[2mm]
&\hspace*{4.2cm}+\,C_F(1-z_p) + \mathcal{I}_{qq}^\text{algo}(z_p)+\delta(1-z_p)I_q^{\mathrm{algo}} \bigg]\,,
\nonumber\\[2mm]
\tilde{j}_{q\to g}\left(z_p,\kappa,\mu_R\right)
&\equiv\frac{\alpha_s(\mu_R)}{2\pi}\left[P_{gq}(z_p)\log\left( \kappa^2(1-z_p)^2\right)+ C_Fz_p 
+ \mathcal{I}_{gq}^\text{algo}(z_p)\right]\,,
\nonumber\\[2mm]
\tilde{j}_{g\to g}\left(z_p,\kappa,\mu_R\right)
&\equiv\delta(1-z_p)+\frac{\alpha_s(\mu_R)}{2\pi}\left[\frac{4C_A(1-z_p+z_p^2)^2}{z_p}
\left(\frac{\log(1-z_p)}{1-z_p}\right)_++P_{gg}(z_p)\log\left(\kappa^2\right)\right.\nn\\[2mm]
& \hspace*{4.2cm}+\, \mathcal{I}_{gg}^\text{algo}(z_p) +\delta(1-z_p)I_g^{\mathrm{algo}}
\bigg]\,,\nonumber\\[2mm]
\tilde{j}_{g\to q}\left(z_p,\kappa,\mu_R\right)&\equiv
\frac{\alpha_s(\mu_R)}{2\pi}\left[
P_{qg}(z_p)\log\left(\kappa^2(1-z_p)^2\right)+ z_p(1-z_p) + \mathcal{I}_{qg}^\text{algo}(z_p)
\right]\;,
\end{align}
where again $\kappa={\cal R}\,p_T^{\mathrm{jet}}/\mu_F''$ and the
$I_{q,g}^\text{algo}$ are as given in Appendix~\ref{appendix:jetfuncs} for the inclusive-jet case. 
To the order we are considering we then have
\begin{align}
{\cal K}_{c\to c'}(z,z_p;\lambda,\kappa,\mu_R)\,=\,\sum_e j_{c\to e}(z,\lambda,\mu_R) \, 
\tilde{j}_{e\to c'}(z_p,\kappa,\mu_R)\;,
\end{align}
and hence from~(\ref{Xsection1})
\beeq\label{Xsection2}
\frac{d\sigma^{H_1H_2\rightarrow (\jet\,h)X}}{dp_T^{\jet} d\eta^\jet d z_h} &= &
\frac{2p_T^{\jet}}{S} \sum_{a,b,c} \int_{x_a^{\mathrm{min}}}^1 \frac{dx_a}{x_a} f_a^{H_1}(x_a,\mu_F)
\int_{x_b^{\mathrm{min}}}^1 \frac{dx_b}{x_b}  f_b^{H_2}(x_b,\mu_F)\nn\\[2mm]
&\times&\int_{z_c^{\mathrm{min}}}^1 \frac{dz_c}{z_c^2} \,
\frac{d\hat{\sigma}_{ab}^c(\hat{s},\hat{p}_T,\hat\eta,\mu_F,\mu_F',\mu_R)}{vdvdw}
\sum_e j_{c\to e}\left(z_c,\frac{{\cal R}\,p_T^{\mathrm{jet}}}{\mu_F'},\mu_R\right)\nn\\[2mm]
&\times&\sum_{c'}\int_{z_h}^1 \frac{dz_p}{z_p}\,
\tilde{j}_{e\to c'}\left(z_p,\frac{{\cal R}\,p_T^{\mathrm{jet}}}{\mu_F''},\mu_R\right)
D_{c'}^h\left(\frac{z_h}{z_p},\mu_F''\right).
\eeeq
In other words, in the NJA the production of a jet with an observed hadron factorizes into the production 
cross section for parton $c$, a jet function $ j_{c\to e}$ describing the formation of a jet ``consisting'' of parton $e$ 
which has taken the fraction $z_c$ of the parent parton's momentum, another jet function $\tilde{j}_{e\to c'}$ 
describing a ``partonic fragmentation'' of parton $e$ to parton $c'$ inside the jet, and finally a regular 
fragmentation function $D_{c'}^h$. This picture is sketched in Fig.~\ref{JetFuncsPic}. It is interesting
to see that the structure of the first two lines of Eq.~(\ref{Xsection2}) is very similar to that of the 
inclusive-jet cross section~(\ref{crsec3}) when formulated in terms of the jet functions ${\cal J}_c$.
In fact, if we drop the last line and perform the sum over parton-type $e$, we will exactly arrive at~(\ref{crsec3}),
since
\beeq
j_{q\to q}\left(z,\lambda,\mu_R\right)+j_{q\to g}\left(z,\lambda,\mu_R\right)&=&{\cal J}_q(z,\lambda,\mu_R)\,,\nn\\[2mm]
2n_f j_{g\to q}\left(z,\lambda,\mu_R\right)+j_{g\to g}\left(z,\lambda,\mu_R\right)&=&{\cal J}_g(z,\lambda,\mu_R)\,.
\eeeq
The last line of~(\ref{Xsection2}) thus describes the production of an identified hadron in the jet. 

We note that at the level of our NLO computation we cannot prove the factorization in~(\ref{Xsection2}) 
to all orders. In fact, at ${\cal O}(\alpha_s)$ we can move terms between $j_{c\to e}$ 
and $\tilde{j}_{e\to c'}$. On the other hand, it seems very natural that the jet functions that we encountered
in the single-inclusive jet case should play a role also in this case in the ``first step'' of the formation of
the final state described by the $j_{c\to e}$. Also, our jet functions $\tilde{j}_{e\to c'}$ are identical to the 
corresponding functions found in the SCET study~\cite{Procura:2011aq} of hadrons in jets produced in 
$e^+e^-$-collisions, except for endpoint contributions $\propto \delta(1-z_p)$ that are necessarily different 
in the SCET formalism due to the presence of a soft function. 

We finally note that the cross section \eqref{Xsection2} may also be expressed in terms of the hadron 
kinematics, using the relation
\begin{align}
\frac{d\sigma^{H_1H_2\rightarrow (\jet\,h)X}}{dp_T d\eta d z_h}\left(p_T,\eta,z_h\right)= \frac{1}{z_h} 
\frac{d\sigma^{H_1H_2\rightarrow (\jet\,h)X}}{dp_T^{\jet} d\eta^\jet d z_h} 
\left(p_T^{\jet} = \frac{p_T}{z_h},\eta,z_h \right)\,.
\end{align}

\section{Phenomenological results \label{pheno}}
We now present some phenomenological results for associated jet-plus-hadron production. 
First, we compare our analytical calculation in the NJA with the one of~\cite{Arleo:2013tya},
where the NLO cross section was obtained numerically by Monte-Carlo integration techniques.
As in that paper, we consider the case of charged hadrons produced in $pp$ collisions at the LHC with
center-of-mass energy $\sqrt{S}=8$ TeV. We define the jet by the anti-$k_t$ algorithm with jet parameter $R=0.4$. 
The renormalization and initial-state factorization scales are set equal to the transverse momentum of the jet,
$\mu_R = \mu_F = p_T^\jet$, while the final-state factorization scale is chosen as $\mu_F^{\prime\prime} = R p_T^\jet$.
The latter choice serves to sum logarithms of $R$ to all orders~\cite{Catani:2013oma,Dasgupta:2014yra},
although this only becomes necessary for jet sizes much smaller than $R=0.4$. 
As in~\cite{Arleo:2013tya} we use the CTEQ6.6M parton distributions~\cite{cteq66} and the 
``de Florian-Sassot-Stratmann'' (DSS07) fragmentation functions of Ref.~\cite{dss07}. 
Our results refer to (summed) charged hadrons, {\it i.e.} $h\equiv h^+ + h^-$.
\begin{figure}[t!]
	\centering
	\includegraphics[width=0.9\textwidth]{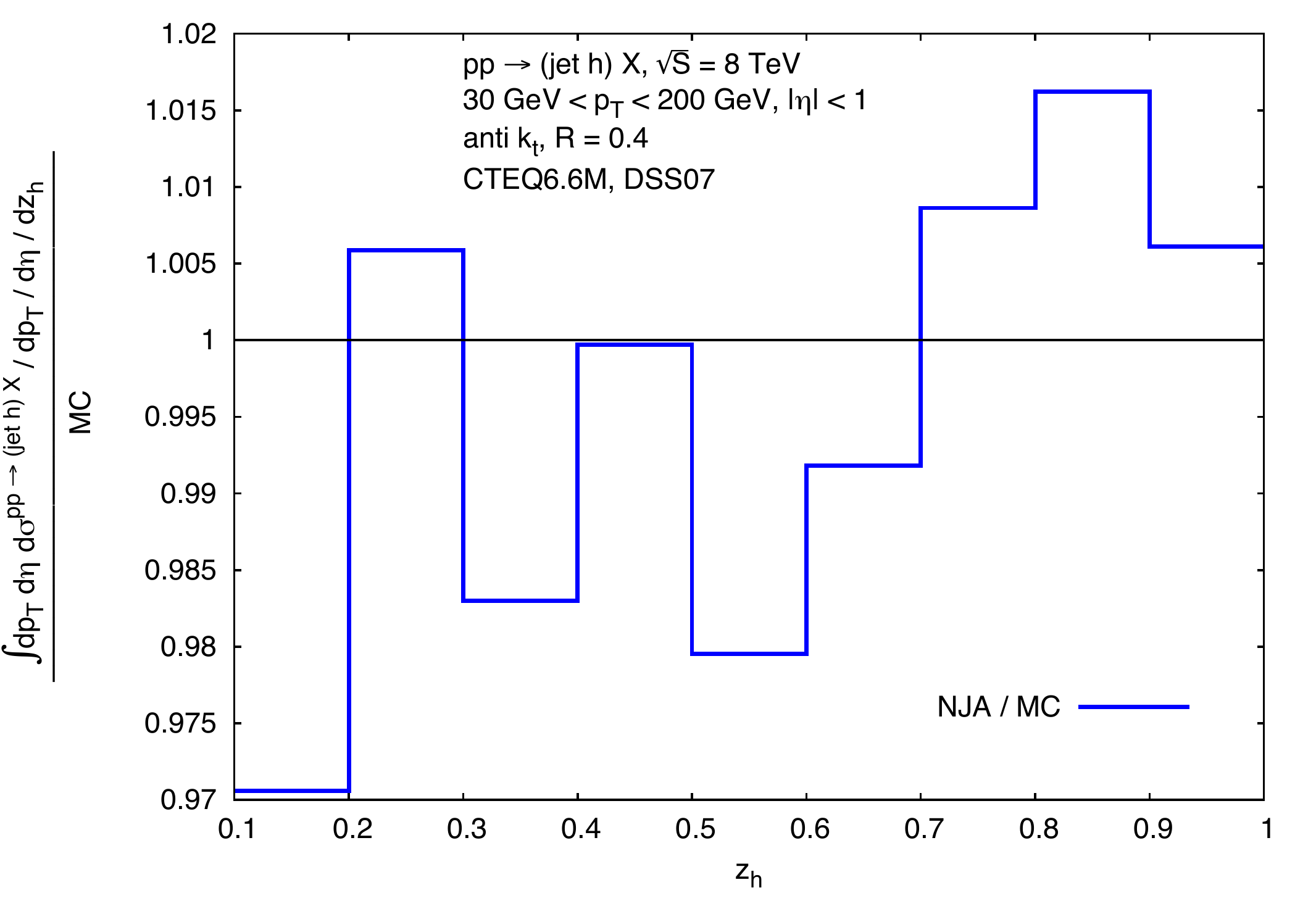}
	\vspace*{-0.3cm}
	\caption{{\sl Comparison of our results in the NJA to the ones of~\cite{Arleo:2013tya} for LHC kinematics.}}
	\label{MCcmp}
\end{figure}

In Fig.~\ref{MCcmp} we show the ratio of the cross section in the NJA with that obtained numerically in
Ref.~\cite{Arleo:2013tya}. The ratio is shown as function of $z_h$, where the cross sections have been 
integrated over $|\eta| < 1$ and 30 GeV $< p_T <$ 200 GeV in hadron rapidity and transverse momentum.
As one can see, the agreement of the two NLO calculations is very good. The deviations are smaller 
than $3\%$ everywhere, which demonstrates the good accuracy of the NJA. We note that in~\cite{Arleo:2013tya} 
a closely related variable $Z_h$ is considered, which is defined as 
\begin{align}
Z_h \equiv \frac{\vec{p}_T \cdot \vec{p}_T^{\;\jet}}{|\vec{p}_T^{\;\jet} |^2}\,.
\end{align}
This definition differs from \eqref{zrel} only by $\mathcal{O}(\mathcal{R}^2)$ corrections, which are anyway
neglected in the NJA. In the limit $z_h\to 1$, the two definitions become equivalent. This explains why the ratio 
in Fig.~\ref{MCcmp} is even closer to unity for larger values of $z_h$. The excellent accuracy of the NJA
observed in the figure is consistent with similar comparisons for the case of single-inclusive jet production 
in the NJA~\cite{Jager:2004jh,Mukherjee:2012uz}. 

Next, we show some results for the kinematics relevant for the ongoing studies in ALICE~\cite{Hess:2014xba}.
We consider $pp$ collisions at $\sqrt{S}=7$ TeV and fragmentation into charged pions ($\pi\equiv \pi^+ + \pi^-$). 
For the rapidity interval we 
choose $|\eta| < 0.5$ and we restrict the jet transverse momentum to 15 GeV $< p_T^\jet <$ 20 GeV. 
As before, the jet is defined by the anti-$k_t$ algorithm with $R=0.4$. We now use more modern sets
for the parton distributions, CT10~\cite{ct10}, and fragmentation functions, DSS14~\cite{dss14}. 
All scales are set equal to the transverse momentum of the jet, $\mu_R = \mu_F = \mu_F^{\prime\prime} = 
p_T^\jet \equiv \mu$.
\begin{figure}[t!]
	\centering
	\includegraphics[width=0.9\textwidth]{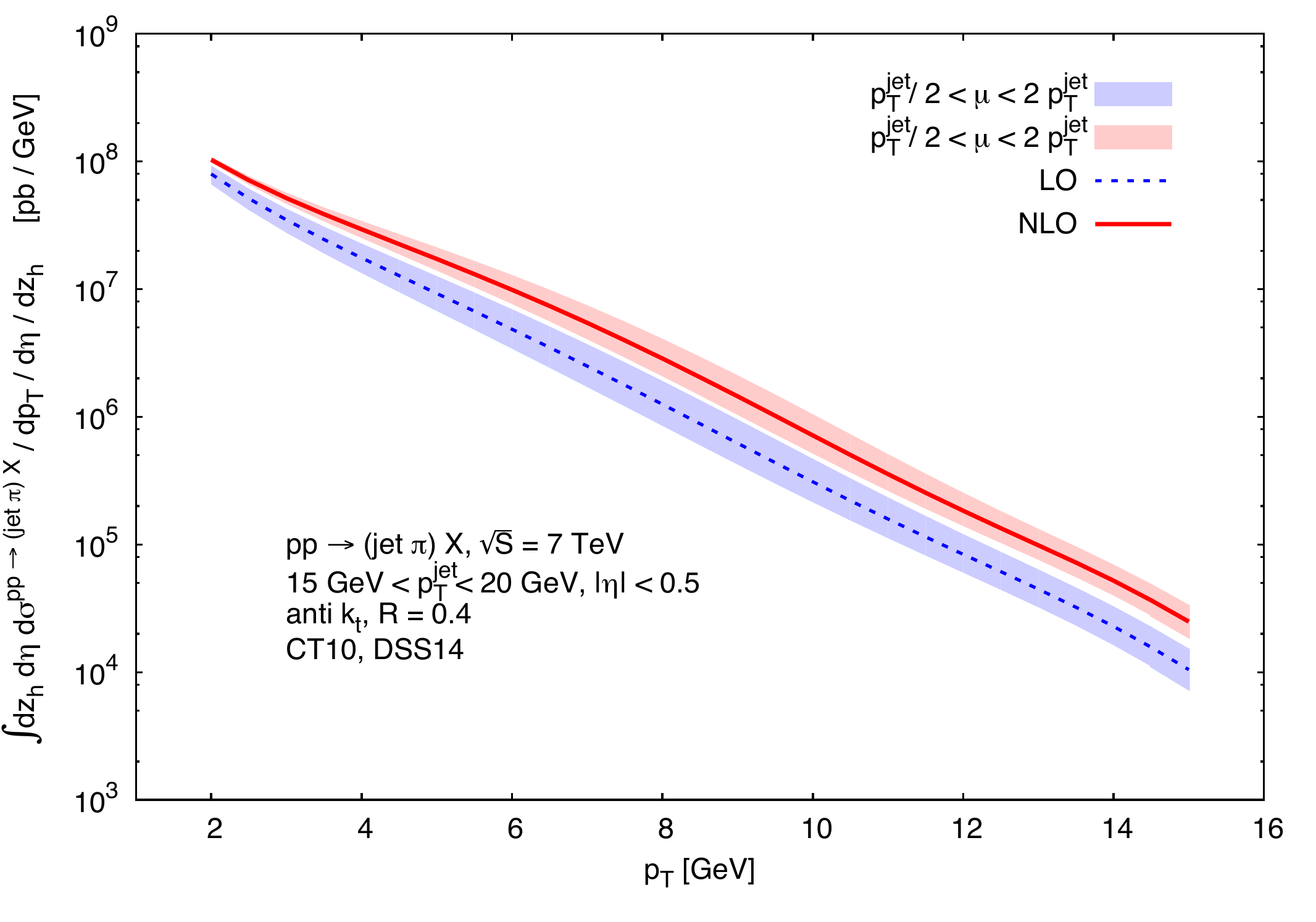}
	\vspace*{-0.5cm}
	\caption{{\sl LO (dashed) and NLO (solid) cross sections for $pp\rightarrow (\jet \,\pi)X$ for ALICE
	conditions, as functions of pion $p_T$. The bands show the scale dependence of the cross section
	for variations of the scale between $p_T^\jet / 2$ (upper end of bands) and $2 p_T^\jet$ (lower end of bands).
	The factorization and renormalization scales have all been set equal and varied simultaneously.}
	\label{diff_pthad}}
\end{figure}
Figure~\ref{diff_pthad} shows the LO (dashed) and NLO (solid) cross sections for associated jet-plus-pion production 
differential in the transverse momentum of the pion. Note that the variable $z_h$ is determined as
$p_T/p_T^\jet$ and hence is varied upon integration over $p_T^\jet$. The bands show the changes of the cross sections 
when the scales are varied in the range $p_T^\jet / 2 < \mu  < 2 p_T^\jet$. As one can see, the scale dependence
of the cross sections improves somewhat when going from LO to NLO, although not as much as one would have hoped. 
This feature was also observed for single-inclusive hadron production in hadronic scattering~\cite{Jager:2002xm}.

\begin{figure}[t!]
	\centering
	\includegraphics[width=0.9\textwidth]{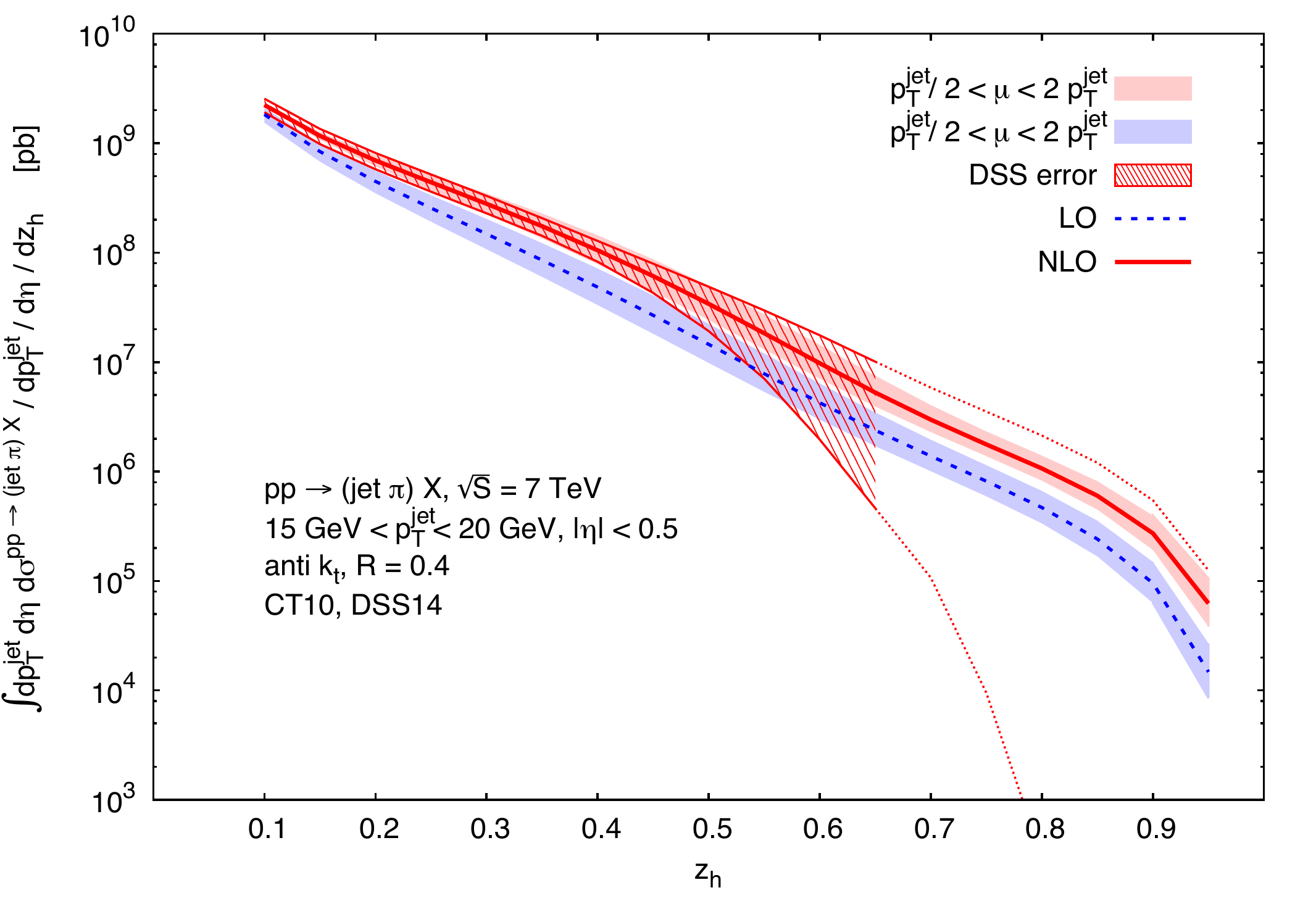}
	\vspace*{-0.5cm}
	\caption{{\sl Same as in Fig.~\ref{diff_pthad}, but as function of $z_h$. As before the solid bands show
	the scale uncertainty. The hatched band displays the uncertainty of the cross section related to the
	fragmentation functions. This band is only reliable up to $z_h=0.65$ and extrapolated beyond (see text).}
	\label{diff_zh}}
\end{figure}

For the same kinematical setup we also show the cross section differential in $z_h$, see Fig.~\ref{diff_zh}. 
A fixed value of $z_h$ implies that the hadron's transverse momentum varies as we integrate over $p_T^\jet$. As
discussed in the Introduction, this arguably is the most interesting distribution for $pp\rightarrow (\jet \,\pi)X$
since it allows direct scans of the fragmentation functions. Apart from the scale variation, we also show in the figure
the uncertainty related to the fragmentation functions, which we compute using the Hessian error sets provided 
in the DSS14 set~\cite{dss14}. Note that the resulting uncertainty band is reliable only up to $z_h \approx 0.65$,
beyond which there are presently hardly any hadron production data available for $e^+e^-$ annihilation or
$ep$ scattering. We hence stop the main uncertainty band there and only sketch its possible extrapolation to 
higher $z_h$. It is clear from the figure that precise measurements of the cross section as a function of
$z_h$ have the potential to provide new information on fragmentation functions that is complementary to --
and in some respects better than -- that available from $e^+e^-$ annihilation. 

\begin{figure}[t!]
	\centering
	\includegraphics[width=0.9\textwidth]{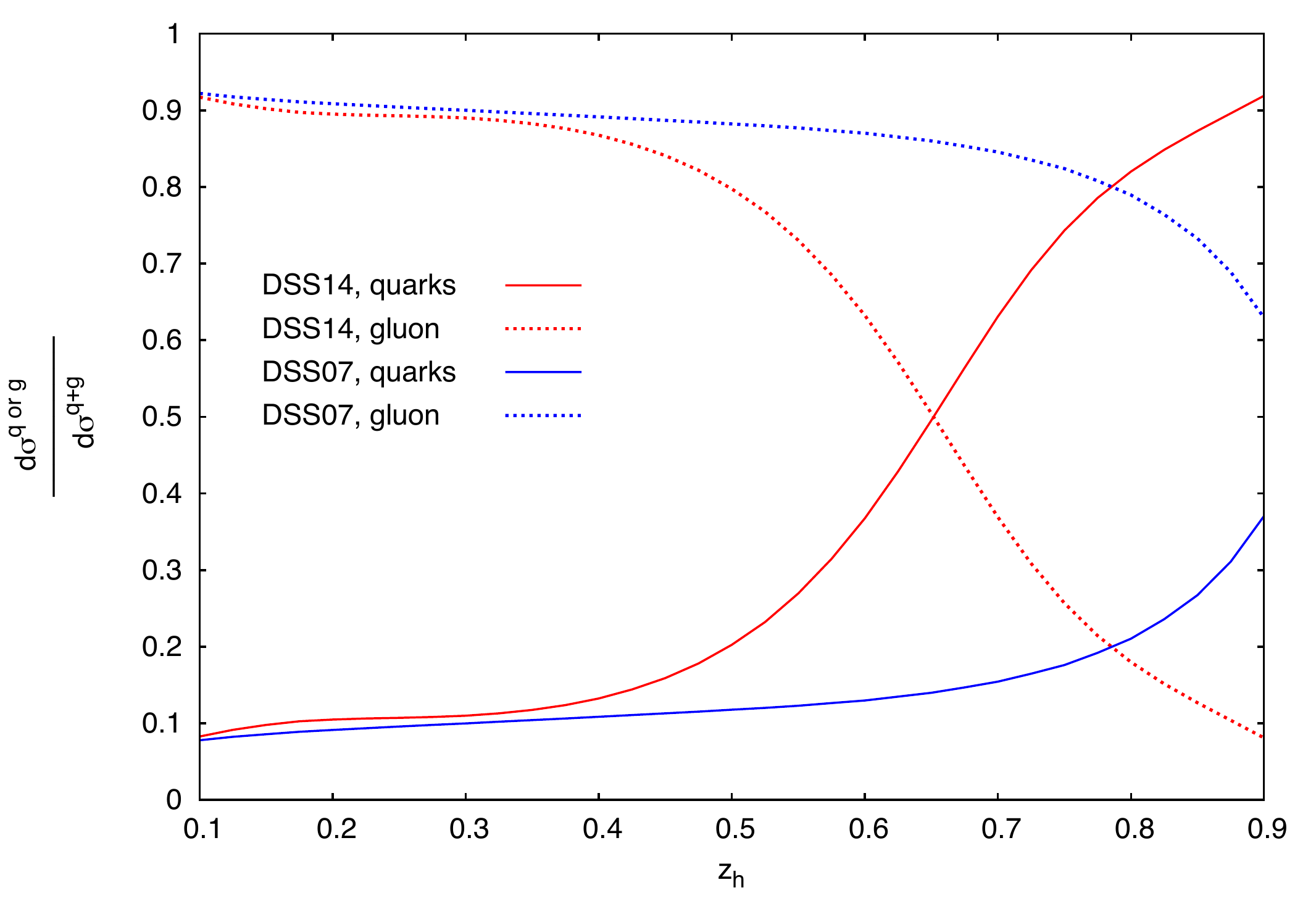}
	\vspace*{-0.5cm}
	\caption{{\sl Normalized quark (solid) and gluon (dashed) contributions to the cross section differential in 
	$z_h$ for the kinematic conditions chosen for Fig.~\ref{diff_zh}. We show results for DSS07~\cite{dss07} 
	and DSS14~\cite{dss14} fragmentation functions.}}
	\label{qg}
\end{figure}

An interesting question is of course which of the fragmentation functions are primarily probed 
when the cross section for $pp\rightarrow (\jet \,\pi)X$ is studied as a function of $z_h$. Depending 
on kinematics, different initial-states may dominate the contributions to the cross section, 
resulting also in different weights with which the fragmentation functions for the various parton
species enter. Given how little information on gluon fragmentation is available from 
$e^+e^-\to hX$ and $ep\to e h X$, it is especially interesting to see how strongly the 
cross section for $pp\rightarrow (\jet \,\pi)X$ depends on $D_g^h$. It is known that for LHC 
energies, channels with gluonic initial states (especially $gg$) typically make important contributions 
to cross sections. In order to explore whether this allows probes of $D_g^h$ at the LHC, we
investigate in Fig.~\ref{qg} the relative contributions of quark/antiquark (summed over all flavors) and 
gluon fragmentation to the cross section for $pp\rightarrow (\jet \,\pi)X$ at ALICE (as shown in the 
previous Fig.~\ref{diff_zh}). We normalize the contributions to the full cross section, so 
that the quark and gluon contributions add up to unity. We use
both the DSS07 and DSS14 sets. As one can see, for $z_h\lesssim 0.5$ the two sets give
similar results and show that the cross section is strongly dominated by gluon fragmentation here.
This is already interesting, since it implies that in this regime clean probes of $D_g^h$ should be
possible that should be much more sensitive than $e^+e^-$ annihilation. Beyond $z_h=0.5$, 
the two sets of fragmentation functions show very different behavior. For DSS07, gluon 
fragmentation continues to dominate all the way up to $z_h\sim 0.9$, whereas for DSS14
the quarks take over at $z_h\sim 0.7$. We stress again that the uncertainties of the fragmentation 
functions become very large at such values of $z$, as we saw in the previous figure, and are
in fact hard to quantify reliably. It is evident that information from $pp\rightarrow (\jet \,\pi)X$
in this regime will be most valuable, regardless of whether quark or gluon fragmentation 
dominates. Detailed measurements for various bins in transverse momentum and rapidity will
likely help in disentangling fragmenting quarks and gluons. 

\begin{figure}[t!]
\vspace*{-2mm}
	\centering
	\includegraphics[width=0.9\textwidth]{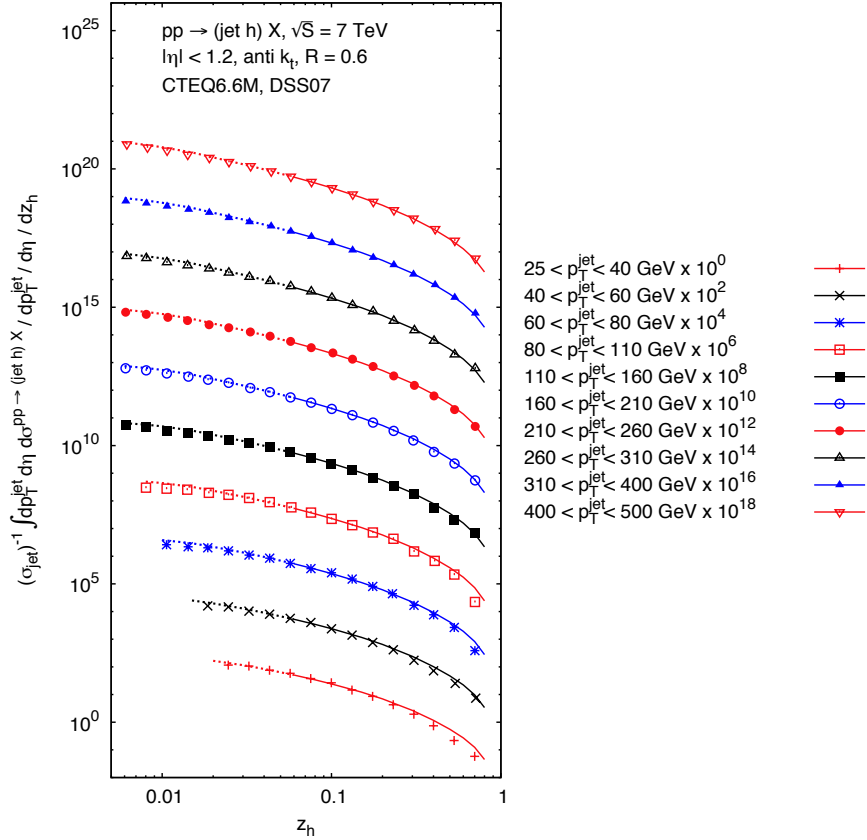}
	\vspace*{-0.6cm}
	\caption{{\sl NLO cross section for $pp\rightarrow (\jet \,h)X$
	as function of $z_h$ at $\sqrt{S} = 7$ TeV, compared to the ATLAS data~\cite{Aad:2011sc}
	for charged hadron production in the leading jet. The cross section is normalized to the total jet rate.
	In the region outside the validity of the DSS07 set the theory curves are extrapolated and plotted as dotted lines.}}
	\label{ATLAS2011}
\end{figure}

\begin{figure}[t!]
\vspace*{-8mm}
	\centering
	\includegraphics[width=0.9\textwidth]{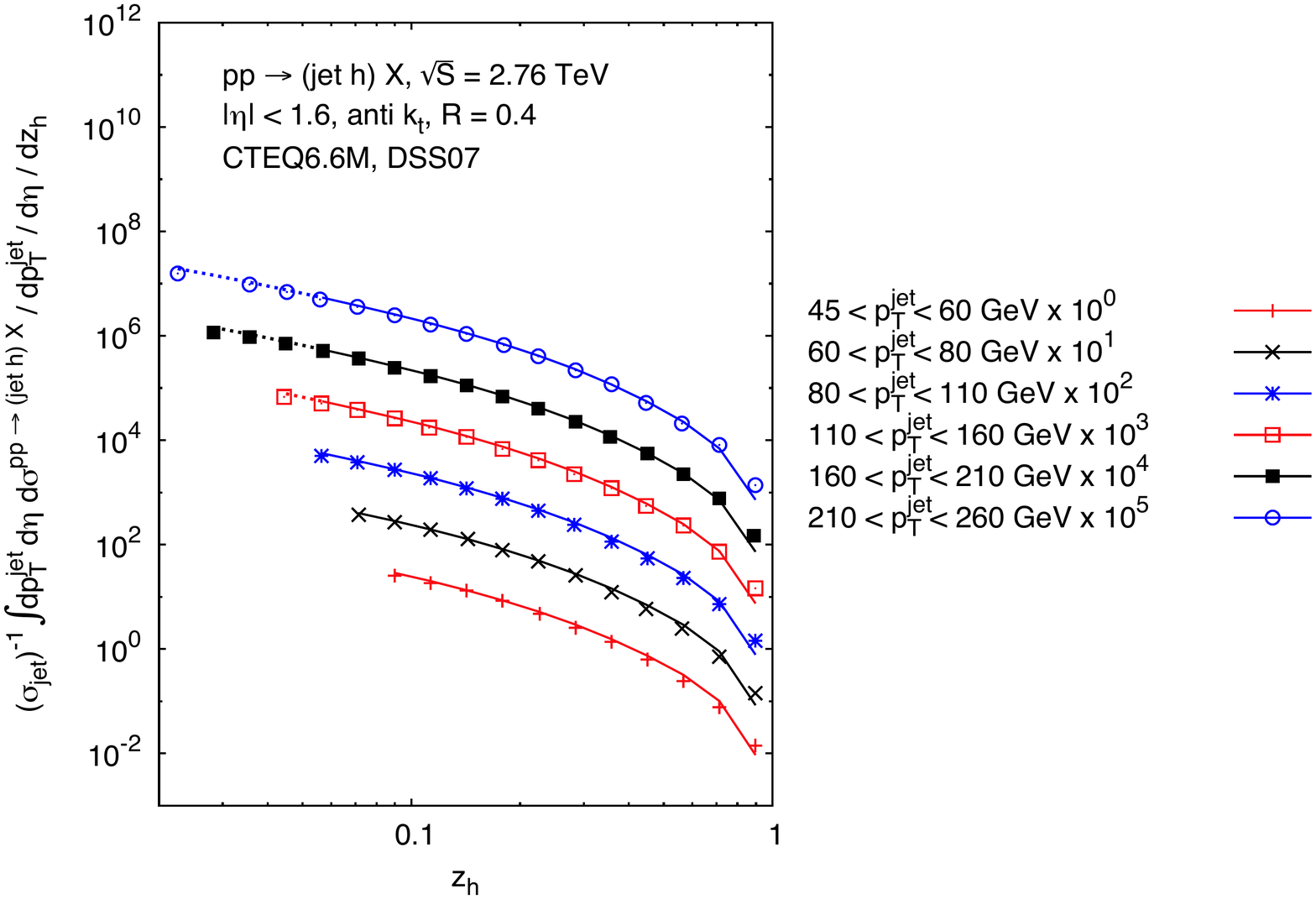}
	\vspace*{-1.2cm}
	\caption{{\sl Same as Fig.~\ref{ATLAS2011} but at $\sqrt{S} = 2.76$ TeV, compared to the preliminary 
	ATLAS data~\cite{ATLAS:2015mla}.}}
	\label{ATLAS2015}
\end{figure}

As mentioned in the Introduction, measurements of charged hadrons produced in jets are already available from 
ATLAS~\cite{Aad:2011sc} and CMS~\cite{Chatrchyan:2012gw}. ATLAS has published measurements 
at $\sqrt{S}=7$~TeV~\cite{Aad:2011sc} and presented preliminary data~\cite{ATLAS:2015mla} 
also at $\sqrt{S} = 2.76$~TeV. The two analyses each use a slightly different definition of $z_h$ 
which however both coincide with our $z_h$ in the NJA limit. Figures~\ref{ATLAS2011} and~\ref{ATLAS2015} 
present comparisons of our NLO calculations to the ATLAS data for the two energies. We have now
gone back to the DSS07 set, since unspecified charged-hadron fragmentation functions 
are not available in the more recent DSS14 set. As one can see, there  is overall a very
good agreement. Note that this agreement extends even down to values of $z_h<0.05$,
well outside the region of validity of the DSS sets. The figures clearly demonstrate the potential 
of the data to further pin down the charged-hadron fragmentation functions.

\begin{figure}[t!]
	\centering
	\includegraphics[width=0.9\textwidth]{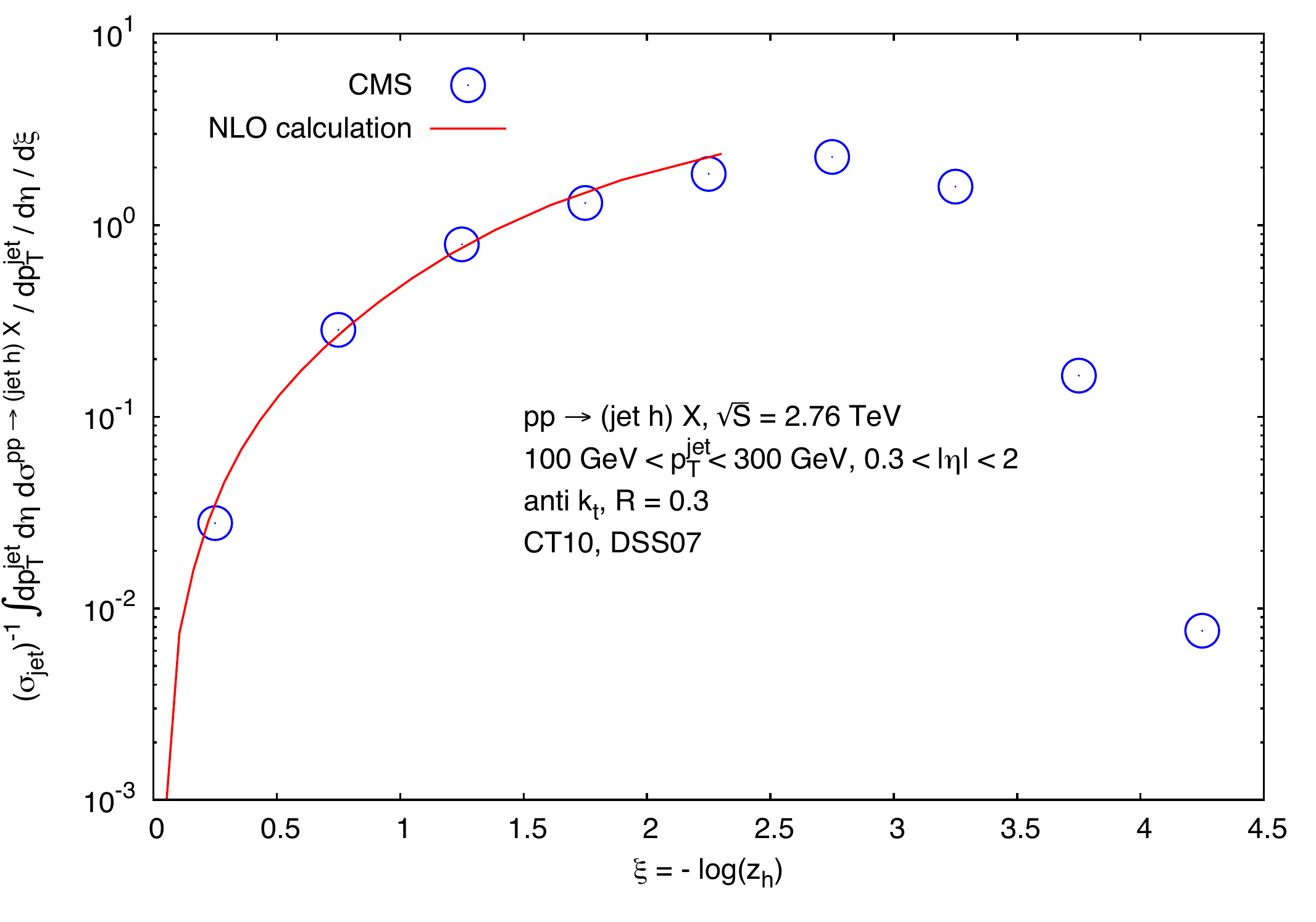}
	\vspace*{-0.5cm}
	\caption{{\sl NLO cross section for $pp\rightarrow (\jet \,h)X$
	differential in $\xi\equiv \log(1/z_h)$ at $\sqrt{S} = 2.76$ TeV, compared to the CMS data~\cite{Chatrchyan:2012gw}
	for hadron production in the leading jet. The cross section is normalized to the total jet rate.}}
	\label{CMS}
\end{figure}

The CMS analysis~\cite{Chatrchyan:2012gw} starts from a dijet sample and then studies charged-hadron 
production inside either the leading jet (which is required to have $p_T^{\mathrm{jet}}>100$~GeV) 
or the subleading jet (with $p_T^{\mathrm{jet}}>40$~GeV). As such, these conditions are different
from the single-inclusive jet situation we consider in this paper, and strictly speaking we cannot
compare to the CMS data. On the other hand, it turns out that the CMS data for hadron production in 
the leading and the subleading jet are in remarkable agreement for $z_h\gtrsim 0.05$, when one
normalizes each of them individually to the corresponding total (leading or subleading) jet event rate. 
This finding clearly indicates that fragmentation inside jets is really independent of the underlying event 
topology and happens in the same way in any jet. Therefore, the overall reservation notwithstanding, we
show in Fig.~\ref{CMS} the comparison of the normalized one-jet rate differential in $\log(1/z_h)$ to the 
CMS data for hadron production in the leading jet. We show the theoretical curve down to $z_h\sim 0.1$.
As one can see, the agreement with the data is very good in this regime. We have found that quark and 
gluon fragmentation contribute roughly in equal parts to the cross section.

\begin{figure}[t!]
	\centering
	\includegraphics[width=0.9\textwidth]{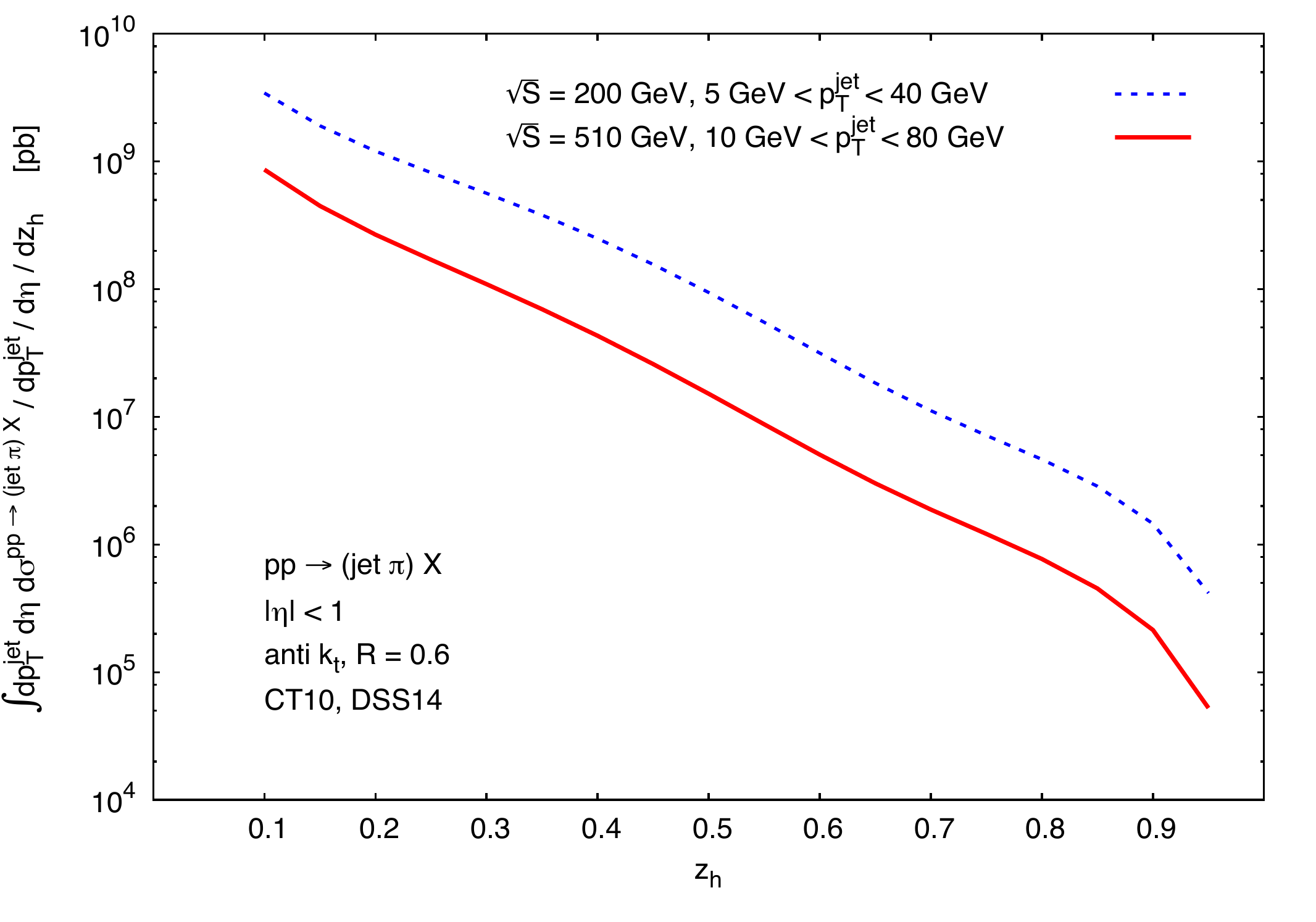}
	\vspace*{-0.5cm}
	\caption{{\sl NLO cross section for $pp\rightarrow (\jet \,h)X$
	differential in $z_h$, for STAR kinematics with $\sqrt{S} = 200 $ GeV (dashed) 
	and $\sqrt{S} = 510$ GeV (solid).}}
	\label{STAR}
\end{figure}

We finally note that measurements of $pp\rightarrow (\jet \,\pi)X$ should readily be feasible at RHIC, 
especially in the STAR experiment where both inclusive jet~\cite{Abelev:2006uq} and pion cross 
sections~\cite{Abelev:2009pb,Agakishiev:2011dc} have been measured. Figure~\ref{STAR} shows
our NLO predictions as functions of $z_h$ for $	pp$ collisions at $\sqrt{S} = 200$ GeV and $\sqrt{S} = 510 $ GeV.
For the former, we have integrated the jet transverse momentum over 5 GeV $< p_T^\jet <$ 40 GeV,
while for $\sqrt{S} = 510 $ GeV we have used 10 GeV $< p_T^\jet < $ 80 GeV. In both cases we integrate
over $|\eta| < 1$. The jet is defined by the anti-$k_t$ algorithm with $R=0.6$. As before we use CT10 and 
DSS14 and set all scales equal to the jet transverse momentum.

\section{Conclusions and outlook \label{Concl}}

We have considered the process $pp\rightarrow (\jet \,h)X$, for which a specific hadron is observed inside a fully 
reconstructed jet. Using the approximation of relatively narrow jets, we have performed an analytical next-to-leading 
order calculation of the partonic cross sections for this process. We have found that the NLO partonic cross 
sections may be systematically formulated in terms of simple jet functions for the process. These functions 
are universal; that is, they only depend on the types of partons producing the jet and fragmenting into the
observed hadron. We note that in the process of computing the jet functions we needed to perform subtractions 
of the final-state collinear singularities. These take the same form as the corresponding subtractions in single-inclusive
hadron production (without a reconstructed jet). This demonstrates that the fragmentation functions are universal 
to NLO in the sense that the same functions appear in $pp\rightarrow (\jet \,h)X$ as in $pp\rightarrow hX$.
Essentially, all effects of the fact that a jet is reconstructed along with the hadron factorize into a perturbatively
computable factor, the jet function. The factorized structure in terms of jet functions we find at NLO suggests 
that this statement is true to all orders. Our finding is in line with the result of~\cite{Procura:2011aq}.

Our numerical results are in very good agreement with those obtained by Monte-Carlo
integration techniques in~\cite{Arleo:2013tya}.
We have presented phenomenological results for the NLO cross section for the kinematics relevant 
for forthcoming measurements at ALICE and for previous ones by ATLAS and CMS. 
These results show that $pp\rightarrow (\jet \,h)X$ 
should enable very sensitive probes of fragmentation functions. In particular, the cross section
differential in $z_h$ probes the fragmentation functions almost ``locally'' at the momentum 
fraction $z_h$. The combination of fragmentation functions that is probed depends on the 
mix of initial-state of parton distributions and hard-scattering functions that dominates. We find
that, in contrast to the standard process $e^+e^-\to hX$ that is customarily used for extractions of
fragmentation functions, the process $pp\rightarrow (\jet \,h)X$ should offer detailed insights into
gluon fragmentation. Also, information at very large $z_h$ might become accessible, although 
here it may become necessary to perform resummations of large logarithmic terms in the
jet functions. We note that at high $z_h$ typical particle multiplicities in the jet a very low, so that
power corrections and non-perturbative phenomena will become important here as well.
As has been discussed in Ref.~\cite{Dasgupta:2007wa}, hadronization corrections to inclusive-jet 
production may exhibit a scaling with $1/R$, making them especially relevant in the case of rather 
narrow jets. Although these corrections are at the same time suppressed by an inverse power 
of transverse momentum, it will be an interesting and important task to investigate their structure
in case of the hadron-plus-jet observable where two separate transverse momenta are present.

There are various other possible extensions of our work that we hope to address in the future.
As is well known, hadron production in jets has important applications in studies of
spin phenomena in QCD in terms of the Collins effect~\cite{Collins:1992kk}, where
the azimuthal distribution of a hadron around the jet axis is considered. 
Studies of the effect in $pp$ scattering~\cite{Yuan:2008yv,Yuan:2008tv,D'Alesio:2010am} 
will require a detailed theoretical understanding 
of the process, to which we hope we have contributed in this paper by computing the 
NLO corrections for the denominator of the spin asymmetry. We expect that our method 
based on jet functions is also applicable to the spin-dependent case. 
Finally, we mention that also photon fragmentation in jets could be interesting as a means to 
constrain the poorly known photon fragmentation functions (see~\cite{Glover:1993xc} for
related work on $e^+e^-$ annihilation and $ep$ scattering).

\section*{Acknowledgments} 

We are grateful to Fran\c{c}ois Arleo, Daniel de Florian, Benjamin He{\ss}, Anne Sickles, 
and Marco Stratmann for useful communications
and discussions. We thank Chi Linh Nguyen for sending numerical results of Ref.~\cite{Arleo:2013tya}.
AM thanks the Alexander von Humboldt Foundation, Germany, for support through a Fellowship for Experienced Researchers.
This work was supported in part by the Institutional Strategy of the University of T\"{u}bingen (DFG, ZUK 63).

\appendix

\renewcommand{\theequation}{A.\arabic{equation}}
\setcounter{equation}{0} 

\section{Details for jet functions in the single-inclusive case}\label{appendix:jetfuncs}

In our results~(\ref{Jinc}) for the single-inclusive jet functions we have the standard LO splitting functions
\begin{align}
P_{qq}(z) &= C_F\left[\frac{1+z^2}{(1-z)_+}+\frac{3}{2}\delta(1-z)\right]\,,\nonumber\\[2mm]
P_{gq}(z) &= C_F\, \frac{1+(1-z)^2}{z}\,,\nonumber\\[2mm]
P_{gg}(z) &= 2 C_A \frac{(1-z+z^2)^2}{z(1-z)_+}+\frac{\beta_0}{2}
\delta(1-z)\,,\nonumber\\[2mm]
P_{qg}(z) &=\frac{1}{2} \left(z^2+(1-z)^2\right)\,,
\end{align}
with $\beta_0 =\frac{11}{3}C_A-\frac{2}{3}n_f$. The ``plus''-distribution is defined as usual by
\beq
\int_0^1 dz \,f(z) [g(z)]_+\,\equiv\,\int_0^1 dz \,(f(z)-f(1)) g(z)\,.
\eeq
Dropping the $\delta$-function contributions and ignoring the ``plus''-distributions, we obtain
the splitting functions $P^<_{ij}(z)$ at $z<1$. For our calculations, we actually need these functions 
computed in dimensional regularization in $D=4-2\varepsilon$ dimensions, where they
are denoted as $\tilde{P}^<_{ij}(z)$. We have
\beq\label{PT1}
\tilde{P}_{ij}^<(z)= P^<_{ij}(z) + \ep P_{ij}^{(\ep)}(z)\,,
\eeq
with 
\beeq\label{PT2}
&&P_{qq}^{(\ep)}(z)\,=\,-C_F(1-z)\,,\quad\quad P_{gq}^{(\ep)}(z)\,=\,-C_Fz\,,\nn\\[2mm]
&&P_{qg}^{(\ep)}(z)\,=\,-z(1-z)\,,\quad\quad P_{gg}^{(\ep)}(z)\,=\,0\,.
\eeeq
We note in passing that the pieces in~(\ref{Jinc}) that are independent of the jet algorithm may 
be constructed following a simple rule: Each of the jet functions ${\cal J}_c$ contains
the combination
\beq
-\frac{\alpha_s}{2\pi}\sum_i\,\left[P_{ij}^<(z) \log\left(\lambda^2(1-z)^2\right) - P_{ij}^{(\epsilon)}(z)\right]\,,
\eeq
up to regularization by distributions at $z=1$.

The algorithm-dependent terms $I_q^{\mathrm{algo}}$ and $I_g^{\mathrm{algo}}$ in Eq.~(\ref{Jinc})
may be determined from the calculations presented in~\cite{Mukherjee:2012uz,Kaufmann:2014nda}.
For cone algorithms we have
\beeq
I_q^{\text{cone}}&=&C_F \left(-\frac{7}{2} + \frac{\pi^2}{3} - 3\log 2\right)\,,\nn\\[2mm]
I_g^{\text{cone}}&=&C_A\left(-\frac{137}{36}+\frac{\pi^2}{3}-\frac{11}{3}\log 2\right)
+\frac{n_f}{2} \left(\frac{23}{18}+\frac{4}{3}\log 2 \right)\,,
\eeeq
while for the (anti-)$k_t$ algorithms
\beeq
I_q^{k_t}&=&C_F \left(-\frac{13}{2} + \frac{2\pi^2}{3} \right)\,,\nn\\[2mm]
I_g^{k_t}&=&C_A\left(-\frac{67}{9}+\frac{2\pi^2}{3}\right)+\frac{23}{18}\,n_f\,.
\eeeq
Finally, for the ``$J_{E_T}$'' algorithm:
\beeq
I_q^{J_{E_T}}&=&C_F \left(-5 + \frac{\pi^2}{2} - \frac{3}{2}\log 2\right)\,,\nn\\[2mm]
I_g^{J_{E_T}}&=&C_A\left(-\frac{45}{8}+\frac{\pi^2}{2}-\frac{11}{6}\log 2\right)
+\frac{n_f}{2} \left(\frac{23}{12}+\frac{2}{3}\log 2 \right)\,.
\eeeq
Interestingly, we find
\beq\label{Irel}
I_j^{J_{E_T}}\,=\,\frac{1}{2}\left(I_j^{\text{cone}}+I_j^{k_t}\right)\,.
\eeq

\section{Jet Functions for the semi-inclusive case}\label{appendix:jetfuncs1}

\renewcommand{\theequation}{B.\arabic{equation}}
\setcounter{equation}{0} 

In addition to $\mathcal{K}^{\mathrm{algo}}_{q\rightarrow q}$ in Eq.~(\ref{Kqq}) we have:
\beeq \label{Kqg}
\mathcal{K}^{\mathrm{algo}}_{q\rightarrow g}\left(z,z_p,\lambda,\kappa,\mu_R\right)&=&
\frac{\alpha_s(\mu_R)}{2\pi}\bigg[-\delta(1-z_p)\Big\{
P_{gq}(z)\log\left( \lambda^2(1-z)^2\right)+ C_Fz\Big\}\nonumber\\[2mm]
&&\hspace*{-3.5cm}+\,\delta(1-z) \Big\{P_{gq}(z_p)\log\left( \kappa^2(1-z_p)^2\right)+ C_F z_p + 
\mathcal{I}_{gq}^\text{algo}(z_p)\Big\}\bigg]\,,
\eeeq
\beeq
&&\hspace*{-2cm}\mathcal{K}^{\mathrm{algo}}_{g\rightarrow g}\left(z,z_p,\lambda,\kappa,
\mu_R\right)\,=\,\delta(1-z)\delta(1-z_p)+
\frac{\alpha_s(\mu_R)}{2\pi}\Bigg[\nonumber\\[2mm]
&&\hspace*{-1.3cm}-\,\delta(1-z_p)\left\{\frac{4C_A(1-z+z^2)^2}{z}
\left(\frac{\log(1-z)}{1-z}\right)_+
+P_{gg}(z)\log\left(\lambda^2\right)
\right\}\nonumber\\[2mm]
&&\hspace*{-1.3cm}+\;\delta(1-z) \left\{ \frac{4C_A(1-z_p+z_p^2)^2}{z_p}
\left.\left(\frac{\log(1-z_p)}{1-z_p}\right)_+
+P_{gg}(z_p)\log\left(\kappa^2\right) + \mathcal{I}_{gg}^\text{algo}(z_p) 
\right\}\right]\,,
\eeeq
\beeq
\mathcal{K}^{\mathrm{algo}}_{g\rightarrow q}\left(z,z_p,\lambda,\kappa,\mu_R\right)&=&
\frac{\alpha_s(\mu_R)}{2\pi}\bigg[-\delta(1-z_p)\Big\{
P_{qg}(z)\log\left(\lambda^2(1-z)^2\right)+ z(1-z)\Big\}\nonumber\\[2mm]
&&\hspace*{-4.2cm}+\;\delta(1-z)\Big\{P_{qg}(z_p)\log\left( \kappa^2(1-z_p)^2\right)+ z_p(1-z_p) + 
\mathcal{I}_{qg}^\text{algo}(z_p) \Big\}\bigg]\,,
\eeeq
where as before $\lambda={\cal R}\,p_T^{\mathrm{jet}}/\mu_F'$, $\kappa={\cal R}\,p_T^{\mathrm{jet}}/\mu_F''$, and
where the algorithm dependent terms are
\begin{align}
\mathcal{I}_{c^\prime c}^\text{algo}(z) = \left\{ \begin{array}{cl} 2 P_{c'c}(z) 
\log\left(\frac{z}{1-z}\right)\Theta(1/2-z) & {\mathrm{cone\;algorithm}}\,,\\[2mm]
2 P_{c'c}(z) \log z & {\mathrm{(anti-)}}k_t {\mathrm{\;algorithm}}\,,\\[2mm]
P_{c'c}(z) \left[\log(z) + \log\left(\frac{z}{1-z}\right)\Theta(1/2-z)\right] & J_{E_T} {\mathrm{\;algorithm}}\,.
\end{array} \right.
\end{align}
We note that a closely related result for the cone algorithm was obtained in~\cite{Procura:2011aq}. 
Again, similar to~(\ref{Irel}), we have
\beq
\mathcal{I}_{c^\prime c}^{J_{E_T}}(z)\,=\,\frac{1}{2}\left(\mathcal{I}_{c^\prime c}^{\text{cone}}+
\mathcal{I}_{c^\prime c}^{k_t}\right)\,.
\eeq

\end{document}